\documentclass{article}
\usepackage{standalone}
\usepackage{amsmath,amssymb}
\usepackage{amsthm}
\usepackage{subcaption}
\usepackage{algorithm}
\usepackage{soul}
\usepackage{algpseudocode}
\UseRawInputEncoding
\algblock{ParFor}{EndParFor}
\algnewcommand\algorithmicparfor{\textbf{parfor}}
\algnewcommand\algorithmicpardo{\textbf{do}}
\algnewcommand\algorithmicendparfor{\textbf{end\ parfor}}
\algrenewtext{ParFor}[1]{\algorithmicparfor\ #1\ \algorithmicpardo}
\algrenewtext{EndParFor}{\algorithmicendparfor}

\usepackage{arxiv}
\usepackage[utf8]{inputenc} 
\usepackage[T1]{fontenc}    
\usepackage{hyperref}       
\usepackage{url}            
\usepackage{booktabs}       
\usepackage{amsfonts}       
\usepackage{nicefrac}       
\usepackage{microtype}      
\usepackage{lipsum}		
\usepackage{graphicx}
\usepackage{wasysym}

\usepackage[
    natbib=true,
    style=numeric,
    sorting=none
]{biblatex}
\hypersetup{
    colorlinks=true,
    linkcolor=blue,
    filecolor=blue,      
    urlcolor=blue
}
\newcommand{\be}{\begin{equation}}
\newcommand{\ee}{\end{equation}}
\newcommand{\bea}{\begin{eqnarray}}
\newcommand{\eea}{\end{eqnarray}}
\newcommand{\bean}{\begin{eqnarray*}}
\newcommand{\eean}{\end{eqnarray*}}

\title{Tracking capelin spawning migration -- Integrating environmental data and Individual-based modeling}
\author{Salah Alrabeei\\
Department of Computer Science\\
Western Norway Univ. of Applied Sciences\\
Bergen, Norway\\
\texttt{salah.alrabeei@hvl.no}\\
\And
Talal Rahman\\
Department of Computer Science\\
Western Norway Univ. of Applied Sciences\\
Bergen, Norway\\
\texttt{talal.rahman@hvl.no}\\
\And
\href{https://orcid.org/0000-0003-0646-9830}{\includegraphics[scale=0.06]{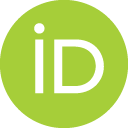}\hspace{1mm}Sam~Subbey}\thanks{Also at Department of Computer Science, Western Norway Univ. of Applied Sciences,	PO Box 7030, 5020 Bergen, Norway}\\
Inst. of Marine Research\\
PO Box 1870\\
5817 Bergen, Norway\\
\texttt{samuels@imr.no}\\
}

\hypersetup{
pdftitle={A template for the arxiv style},
pdfsubject={q-bio.NC, q-bio.QM},
pdfauthor={David S.~Hippocampus, Elias D.~Striatum},
pdfkeywords={First keyword, Second keyword, More},
}
\begin{document}
\maketitle
\begin{abstract}
This paper presents a modeling framework for tracking the spawning migration of the capelin, which is a fish species in the Barents Sea. The framework combines an individual-based model (IBM) with artificial neural networks (ANNs). The ANNs determine the direction of the fish's movement based on local environmental information, while a genetic algorithm and fitness function assess the suitability of the proposed directions. The framework's efficacy is demonstrated by comparing the spatial distributions of modeled and empirical potential spawners.

The proposed model successfully replicates the southeastward movement of capelin during their spawning migration, accurately capturing the distribution of spawning fish over historical spawning sites along the eastern coast of northern Norway.

Furthermore, the paper compares three migration models: passive swimmers, taxis movement based on temperature gradients, and restricted-area search, along with our proposed approach. The results reveal that our approach outperforms the other models in mimicking the migration pattern. Most spawning stocks managed to reach the spawning sites, unlike the other models where water currents played a significant role in pushing the fish away from the coast. The temperature gradient detection model and restricted-area search model are found to be inadequate for accurately simulating capelin spawning migration in the Barents Sea due to complex oceanographic conditions.
\end{abstract}
\section{Introduction}
\begin{figure}[!ht]
\centering
\includegraphics[scale=0.5]{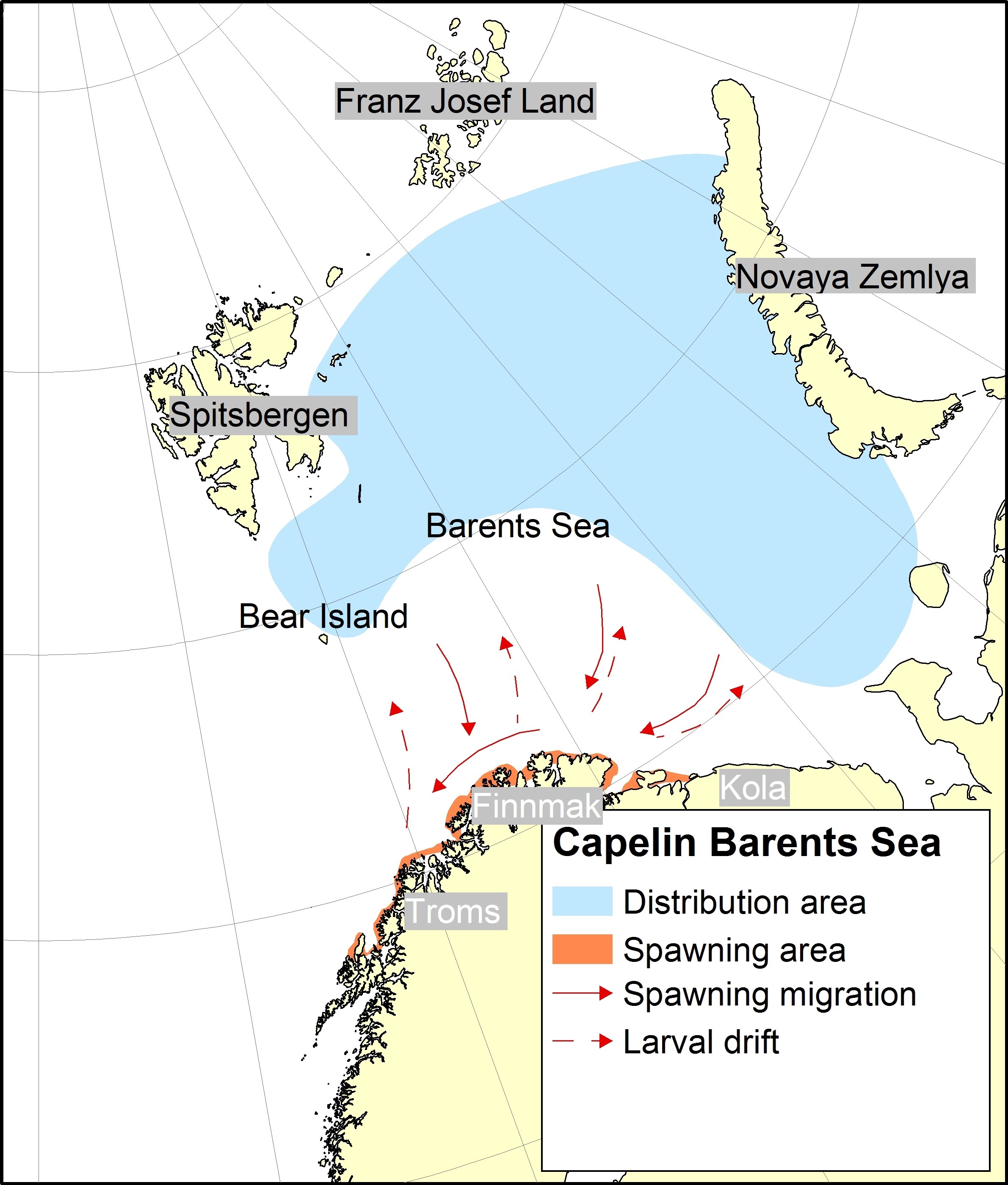}
\caption{ Capelin grounds in the Barents Sea. Used with courtesy of the Institute
of Marine Research, Norway}
\label{fig:Capelin_areas}
\end{figure}

The Barents Sea capelin, hereafter referred to as capelin, is a small pelagic fish that inhabits the Barents Sea at all stages of its life \cite{gjosaeter1998population, carscadden2013recruitment}. Capelin exhibits schooling behavior and undergoes extensive seasonal migrations throughout its life cycle. During winter and early spring, mature capelin migrate from their wintering habitat in the central Barents Sea toward the spawning grounds near the coasts of northern Norway (Troms and Finnmark counties) and Russia (Kola Peninsula) \cite{christiansen2008facultative} (see Figure \ref{fig:Capelin_areas}). 

The literature is limited regarding factors that influence capelin migration routes and final destinations (spawning locations) \cite{carscadden1997effects, carscadden2013comparison}. The previous conclusion of studies (\cite{ozhigin1985some, huse2008capelin}) suggesting that water temperature may be the central factor has been contradicted by recent findings in \cite{alrabeei2021spatial}. Therefore, it is plausible that multiple environmental factors act in concert to influence the migratory behavior of this spawning species. Providing an explicit description of how these various factors may interact to dictate the spawning migration is challenging. This is because both the interactions and their combined effect on the spawning migration are nonlinear.

Tracking migration routes for marine species is generally challenging. However, modeling feeding migration routes, including those of capelin, may benefit from the incorporation of knowledge about the spatial-temporal distribution of food densities into the modeling framework \cite{huse2008capelin}. The process of modeling spawning migration routes may be simplified by a priori ecological knowledge, such as the propensity to migrate along the coast to well-defined spawning sites, using currents for passive transport \cite{tu2012using}. In the case of capelin, spawning migration routes do not align with the coast but cover several hundreds of kilometers against water currents \cite{gjosaeter1998population}. 

We focus on spawning migratory behavior because capelin migrate against the current for hundreds of kilometers \cite{gjosaeter1998population}, unlike many other species in various environments that follow the coasts \cite{gjosaeter1998population}. Furthermore, during spawning migration, capelin do not follow food density, as their goal is to reach the spawning region and choose the best spawning site \cite{gjosaeter2016predation, ozhigin1985some}. Therefore, modeling spawning migration is more challenging since capelin may not orient themselves based on environmental factors such as food density when moving. Additionally, spawning migration patterns vary between years, possibly in response to climate change \cite{huse2008capelin}.

Individual-based (also known as agent-based) models (IBMs) aim to capture the continuously changing rules underlying individual animal behavior \cite{huse2002modelling, deangelis2018decision}. For instance, IBMs have been used to investigate how individual species respond to external environmental stimuli \cite{byron2014salmon} and to understand emergent collective behavior \cite{vicsek2012collective, cichos2020machine}. According to \cite{deangelis2005individual}, the IBM approach surpasses classical models in five different categories when studying variability in space, lifecycle details, phenotype trait variations and behavior, learning and experience, and genetics and evolution.

Developing a realistic IBM requires high data throughput to accurately capture fine-scale individual characteristics \cite{watkins2013evaluating}. However, improved computing resources facilitate the application of IBMs to model high-resolution spatial dynamics of animals \cite{wang2005parallel}.

A major challenge in the application of IBMs is the difficulty of identifying and integrating crucial underlying behavioral rules into the model definition \cite{bayindir2016review, trianni2008evolutionary}. Complex systems, such as fish and birds, have evolved sensing capabilities that allow them to gather information about their environments or communicate \cite{yeomans2017nature}. Individuals within a school of fish or a flock of birds exchange information as part of a process to self-organize into a collective state \cite{attanasi2014information}. The fundamental challenge posed by the high degrees of freedom in complex systems is not limited to IBMs alone but applies to all mechanistic models \cite{cichos2020machine}.

Machine learning tools have recently become valuable in advancing our understanding of biological systems \cite{floreano2008bio}. Progress in this area has been accelerated by the remarkable success of neural network-based algorithms in classification and image recognition \cite{lecun2015deep}. The classical neural network consists of a series of algorithms designed to recognize underlying relationships in a set of data, using a procedure that mimics the operational workings of the human brain \cite{anthony1999neural}.
 
This paper presents a framework for effectively modeling the spatio-temporal behavior of capelin in the Barents Sea during their spawning migration. Our approach incorporates an adaptive agent-based framework that considers capelin's environmental conditions and decision-making processes.

In this study, our focus is on reproducing the specific spawning migration of the Barents Sea capelin during the winter of 2019. Our primary objectives are to characterize the pattern of spawning migration, analyze the timing of spawning, and examine the selection of spawning sites. To ensure the robustness of our framework, we conducted sensitivity tests on various sub-models. Furthermore, we compared the outcomes of our model with those of traditional models and validated our findings using observational data. The structure of the paper is as follows: Related work is discussed in section \textbf{\ref{sec.2}}. Then, our computational framework is introduced in section \textbf{\ref{sec.3}}. Section \textbf{\ref{sec.4}} discusses how to simulate capelin spawning migration and the model sensitivity. The simulation results and discussions are presented in section \textbf{\ref{sec.5}}.
\section{Previous and related work}
\label{sec.2}
Due to the limited understanding of fish-to-fish and fish-to-environment interactions, there is a lack of definitive model assumptions that accurately capture fish behaviors and navigation mechanisms toward their destinations. Consequently, various models have been devised to depict fish migration and population dynamics. At the population level, continuum-based models describe average quantities, including mean density and mean velocity, about the entire population \cite{degond2018mathematical, carrillo2017particle, smith2018spatially}.

In particular, Individual-Based Models (IBMs) are frequently utilized to depict trajectories, interactions among individuals within a school, and their responses to the surrounding environment \cite{smith2018spatially}. Considering that fish swimming behavior varies significantly depending on factors such as species, habitat, and the purpose of movement (e.g., migration, schooling, escaping), existing fish migration models primarily differ in terms of how fish movement is defined and what factors govern their velocity function $F$ in ( \ref{eq.1}):
\begin{equation}
    \vec{X}(t+\Delta t) = \vec{X}(t) + \int_{t} ^ {t+\Delta t} F(\vec{X},\Theta(\vec{X},\tau),\tau) d\tau.
    \label{eq.1}
\end{equation}
Here, $\vec{X(t)} \in \mathbb{R}^{2}$ represents the position of the individual fish at time $t$, and $F(\vec{X},\Theta,\tau)$ is the velocity function of fish, dependent on the temporal position $\vec{X}$ and a function of environmental variables $\Theta(\vec{X},\tau)$.

The function $F$ in (\ref{eq.1}) is defined in various ways, depending on the assumed level of knowledge possessed by the fish. Some models consider environmental factors, such as temperature \cite{barbaro2009modelling, tu2012using, hubbard2004model}, prey density \cite{wang2013coupling}, or bathymetry \cite{politikos2015simulating}, as sources of information that influence the direction of fish movement. Moreover, these models differ in the movement mechanisms governing how individual fish navigate toward preferred habitats based on their assumed knowledge of the environment. Two commonly used movement mechanisms are \textit{gradient detection} and "local search," both of which have been applied in several fish migration models (see \cite{watkins2013evaluating}). The sea surface temperature (SST) is used as the movement-controlling variable in both movement mechanisms.

\subsubsection*{Gradient detection}
In the gradient detection approach, fish are assumed not to have direct knowledge of the surrounding temperature but can detect temperature gradients \cite{tu2012using, politikos2015coupled, barbaro2009modelling}. In this approach, the eastward and northward swimming speeds $Vx_{k}$ and $Vy_{k}$ are given by:

\begin{equation}
    F(\vec{X},\Theta,\tau) = \frac{\nabla(T(\vec{X},\tau))}{|| T(\vec{X},\tau)||} + V(\vec{X},\tau).
    \label{eq.gradTemp}
\end{equation}

Here, $T$ and $V$ represent the sea surface temperature (SST) and water velocity at position $\vec{X}$ and time $t$, respectively.

\subsubsection*{Local-search movement}
In the local search movement approach, fish are assumed to assess their surroundings and move towards the habitat with the highest temperature, representing their preference \cite{xu2013environmental, politikos2015simulating}. In practice, fish evaluate all neighboring cells within their swimming distance and select the one with the highest temperature as the optimal location. The velocity function in (\textbf{Eq.} \ref{eq.1}) is given by:

\begin{equation}
 F(\vec{X},\Theta,\tau) = \bigg[ \cos{\theta} ,  \sin{\theta} \bigg],
\end{equation}
where $\theta = \arccos \bigg( \frac{\vec{X(t)} \cdot \vec{X}_{optT}(t)}{||\vec{X}(t) || \hspace{3dd} ||\vec{X}_{optT}(t) ||}\bigg)$, and $\vec{X}_{optT}$ is the location of the optimal (maximum) SST surrounding the fish.

While the movement of fish in these approaches is controlled by one or more environmental variables such as temperature and food density \cite{barbaro2009modelling, tu2012using, politikos2015simulating}, explicit rule-based models may not be feasible when the control variables are unknown to a specific fish species in a particular habitat or when the velocity function (see \textbf{Eq.} \ref{eq.1}) involves nonlinearity and complex relationships. In such cases, machine learning techniques provide a flexible strategy that can replace fixed rules governing fish movement decisions \cite{huse1998ecology}. Machine learning-based approaches have been commonly used for modeling fish behaviors over the past four decades, with various algorithms employed to capture spatial memory and behavior based on available species and environmental data \cite{cichos2020machine}. Artificial neural networks (ANNs), a type of machine learning method initially designed to simulate brain activity, have been utilized extensively. ANNs consist of virtual neurons arranged in layers, connected by weighted links representing synapses \cite{deangelis2018decision, bennett2006modelling}.

Artificial neural networks (ANNs) learn by adjusting their weights to process inputs and generate appropriate outputs \cite{huse1998ecology}. Learning in ANNs typically requires data to optimize and update the weights. However, acquiring such data may not always be feasible in fish behavioral modeling. In such cases, heuristic optimization methods like genetic algorithms (GAs) can be employed to train ANN-based movement decisions without predefined data \cite{dagorn1997simulation, huse1998ecology, huse2001modelling, bennett2006modelling, okunishi2009simulation}. The genetic algorithm (GA) is a heuristic technique that applies evolutionary principles such as crossover, mutation, and natural selection to find optimal solutions to problems \cite{holland1975adaptation, deangelis2018decision}.

In this study, we employ evolutionary artificial neural networks as the adaptive decision-making sub-model within the Lagrangian Individual-Based Model (IBM) framework to simulate the spawning migration behaviors of capelin. Unlike traditional approaches that require predefined migration routes for training, our method allows capelin to adapt its decision-making processes dynamically during the migration. By combining evolutionary artificial neural networks with the Lagrangian IBM, we aim to capture the complex and dynamic nature of capelin spawning migration without relying on predetermined routes.
\section{Computational Framework}
\label{sec.3}
In this framework, we employ evolutionary artificial neural networks as the adaptive decision-making sub-model within the Lagrangian Individual-Based Model (IBM) to simulate the spawning migration behaviors of capelin. Unlike traditional approaches that require predefined migration routes for training, our method allows capelin to adapt its decision-making processes dynamically during the migration. By combining evolutionary artificial neural networks with the Lagrangian IBM, we aim to capture the complex and dynamic nature of capelin spawning migration without relying on predetermined routes. The mature capelin population is modeled as an autonomous purposeful agent that interacts with the surrounding environment, representing the Barents Sea.
\begin{figure}[!ht]
    \centering
\includegraphics[width=0.8\linewidth]{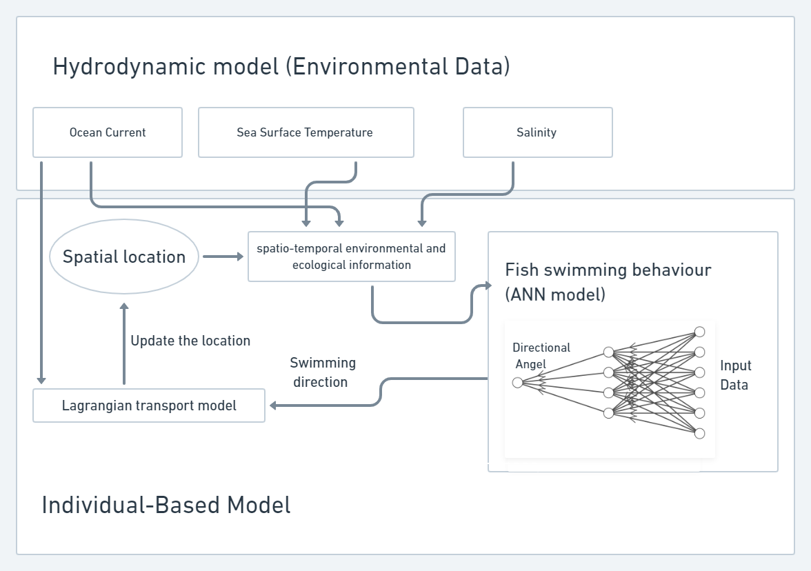} 
\caption{Flowchart of the Adaptive migration model structure.} \label{fig:Framework}
\end{figure}
\subsection{Simulation environment}
The capelin spawning migration has historically occurred within a spatial range defined by the latitude range of $69-80^\circ$N and the longitude range of $6-53^\circ$E \cite{gjosaeter1998population, ozhigin1985some}. Our model represents this bounded spatial domain by discretizing it into an equally spaced mesh of $564 \times 133$ nodes. The horizontal spatial resolution of the mesh was ${\frac{1}{12}}^\circ$ (approximately 9.26 km).

Observations and previous studies have indicated that satellite sea surface temperature (SST) \cite{huse2008capelin, ozhigin1985some}, ocean surface currents (OSC) \cite{gjosaeter1998population}, and water salinity (WS) \cite{Magnsson20051AD} play a role in influencing the spawning migration pattern of capelin. To incorporate these environmental factors into our model, we defined three digital environmental fields: SST, OSC, and WS. The data for these fields were obtained from the freely available Copernicus Marine Environment Monitoring Service (CMEMS). The spatial horizontal resolution of the data is provided in regular grids of size $1/12^{\circ}$ (approximately 8 km). In contrast, the vertical resolution is irregular, with increasing distances between layers in deeper waters. The temporal resolution of the data is one day (1d). For simplicity, we ignored daily vertical movement during the horizontal migration and averaged the data for each environmental field at the first 30 meters depth to represent a single layer. The data for the three digital fields were stored in the main model mesh, with each grid point storing values for SST, OSC, and WS.
\subsection{Capelin agents}
Each capelin agent ($k$) is represented by an array of variables that describe its spatial locations ($X_k$, $Y_k$), swimming velocities ($Vx_{k}$, $Vy_{k}$), decision-making processes, and information on arrival time and travel duration. Each agent is assumed to store partial spatial information about its environment to make informed temporal movement decisions. This information includes variables such as the current state, surrounding temperature, currents, distance to the destination, and other relevant factors that guide movement decisions at each time step.

Since mature migrating capelin can vary in body length ($bl$) between 14 and 20 cm \cite{gjosaeter2016predation}, and observed capelin swimming speeds fall within the range of $1.5$ to $1.7$ body lengths per second \cite{behrens2006swimming}, the capelin agents are assumed to have different swimming speeds ranging from $0.21$ to $0.34$ m/s. To ensure that there is no loss of information transfer between the environment and the agent during the simulation, considering the potential imbalance between spatial and temporal resolutions, the temporal resolution is set to 6 hours. This choice satisfies the Courant-Friedrichs-Lewy (CFL) criterion \cite{martin1993vegetation}:

\begin{equation*}
    \delta t \le \frac{\delta \bar X}{s_{max}}
\end{equation*}

Here, $\delta t$ is the time step, $\delta \bar X$ is the spatial step, and $s_{\text{max}}$ is the maximum swimming speed of the capelin agents. Given a maximum swimming speed of 1.22 km/h and a spatial resolution of 9.26 km, the acceptable temporal resolution is $\delta t = 6$ hours, which satisfies $\delta t \leq 9.26$ km / 1.22 km/h $\approx 7.56$ hours.

\subsection{Movement}
The movement of the capelin agents is determined by a two-dimensional Lagrangian transport model governed by Equation (\ref{eq.1}). At each time step, the position of the agent is updated based on the velocity function $F(\vec{X},\Theta,t)$, defined as:

\begin{equation}
F(\vec{X},\Theta,t) = F_v(\vec{X},\Theta,t) + F_{adv}(\vec{X},t) + \lambda \sqrt{\delta t \cdot D}
\end{equation}

Here, $F_v$ represents the fish swimming velocity, $F_{adv}$ represents the seawater current velocity and the last term accounts for dispersion due to turbulence. The dispersion term is modeled using a normally distributed random number $\lambda$ ranging from -1 to +1, and $D$ is the horizontal dispersion coefficient, set to $100$ m$^2$/s \cite{adlandsvik1994modelling}).

The fish swimming velocity $F_v$ is determined by three fully connected feed-forward artificial neural networks (described in detail in the next subsection). These networks take several environmental and ecological variables as inputs and produce the swimming velocity as output.

\subsection{Learning and Adaptation}
Our model differs from previous models in terms of the knowledge and learning capabilities of the capelin agents and their use in guiding movement. We assume capelin agents store information about their local spatial and temporal environments to make informed movement decisions. The knowledge about the environment is bounded by the maximum rate of movement per day, spatially, and by one day, temporally. The capelin agents in our model learn to replicate the migratory behavior of real capelin fish by making decisions based on learned stimulus-response relationships with the environment. Each agent is equipped with evolutionary artificial neural networks that optimize their fitness function to facilitate learning.

\subsubsection{Artificial Neural Network (ANN)}
We propose that each capelin agent determines its directional movement based on a nonlinear function of various environmental and ecological variables, utilizing its bounded knowledge of the environment. This function is decomposed into three fully connected feed-forward artificial neural networks. The input layer of the ANN receives relevant information for decision-making, while the outer layer produces the directional angle \cite{behrens2006swimming}. The first two nodes of the input layer receive sea surface temperature (SST) data for the current day and the day before, respectively. The third and fourth nodes receive salinity and current magnitude information. The fifth node represents the distance to the spawning region, and the sixth node represents the longitudinal degree. The input layer is connected to the hidden layer through weighted connections, determining the strength of the connections. The hidden layer is then connected to the output layer, determining the directional angle. The values are biased and passed through the standard Sigmoid function for conversion \cite{anderson1995introduction}. Each node in the output layer corresponds to one of the eight directional angles, representing angles of $\frac{j\pi}{4}$, where $j$ ranges from 1 to 8.

Each capelin agent provides spatio-temporal information to its ANN at each time step and receives the corresponding directional angle. Each ANN consists of three layers: an input layer with six nodes, a hidden layer with 18 nodes, and an output layer with eight nodes representing the directional angles. Thus, each ANN contains a total of 252 nodes.

Mathematically, the nonlinear function represented by the ANN connections can be expressed as follows:

\begin{equation}
\begin{split}
    O_1 = W_{ih}^T I_k + b_1 \hspace{10dd}\\
    TO_1 = \sigma ( O_1) = \frac{1}{1+ e^{- O_1}} \\
    O_2 = W_{ho}^T TO_1 + b_2 \\
    Y_k  = \sigma ( O_2) = \frac{1}{1+e^{- O_2}} 
\end{split} 
\label{system}
\end{equation}
where $\sigma$ represents the sigmoid function, as described in \cite{anderson1995introduction}. The input and output arrays of each individual $k$ are denoted as $I_k$ and $Y_k$, respectively. The random weight matrices connecting the input and hidden layers ($W_{ih}$) and the hidden and output layers ($W_{ho}$) are represented as well. Additionally, $b_1$ and $b_2$ are arrays of random values.

The system defined in Equation (\ref{system}) can be expressed concisely as:

\begin{equation}
Y_k = \sigma \left( W_{ho}^T \sigma \left( W_{ih}^T I_k + b_1 \right) + b_2 \right).
\end{equation}

All values in the $I_k$ array are standardized to fall between 0 and 1. The output $Y_k$ is an array of length 8, with the maximum value indicating the directional angle $\theta_k$:

\begin{equation}
\theta_k = \frac{\pi}{4} \Psi_k \left( \max(Y_k) \right),
\end{equation}

where $\Psi_k: Y_k \rightarrow Z_k$ is a bijective operator. $\Psi_k(y_j) = j$ for $j = 1,2,..., |Y_k|$. In simpler terms, $\Psi_k$ is a function that returns the index of the maximum value in the output array.

\subsubsection{Evolutionary adaptation}  

This work does not consider direct local interactions between agents during migration. Instead, an evolutionary adaptation process inspired by the genetic algorithm is developed parallelly (see pseudo-code in Algorithm \ref{alg:cap}). Multiple releasing points are defined to simulate the natural spawning migration of capelin fish, where thousands of agents are initially released. It is acknowledged that the migratory behaviors of capelin fish are not necessarily identical due to their diverse spatial locations and varying environmental conditions. Therefore, the adaptation process in our model is performed in parallel, with separate genetic algorithms running for each sub-population. Each sub-population is identified based on its origin (releasing point), ensuring that evolutionary migration occurs within populations with similar characteristics and habitat preferences.

In practice, the genetic algorithm is employed to train the artificial neural networks (ANNs) responsible for generating movement decisions. The initial weights of each ANN, corresponding to each agent fish in the first and second layers, are randomly generated from a truncated normal distribution. The mean of the distribution is zero, and the standard deviation is calculated as $ \sqrt{\frac{2}{ N_i + N_h}}$ and $ \sqrt{\frac{2}{ N_{h} + N_{o}}}$ for the first and second layers, respectively. Here, $N_i$, $N_h$, and $N_o$ represent the number of nodes in the input, hidden, and output layers, respectively.

\subsubsection*{Evaluation}
In the context of genetic algorithms applied to behavioral ecology, defining a realistic fitness function that accurately evaluates fish behaviors can be challenging \cite{calvez2005automatic}. In this study, the fitness function aims to assess how well the simulated capelin fish can safely cover hundreds of kilometers within a specific timeframe to reach their spawning sites. The fitness function consists of the swimming fitness component ($f_d$) and the temperature fitness component ($f_t$), both weighted to minimize the distance to spawning sites and reduce predating risk. The fitness function of agent $j$ is computed as a weighted sum of these two components, as shown in Equation (\ref{eq:fitness_fun}).
\begin{equation}
f_j = \alpha f_d^{-1} + (1-\alpha)f_t.
\label{eq:fitness_fun}
\end{equation}
where $\alpha$ is a weighting factor that controls the relative influence of the two fitness components. It is computed dynamically during the simulation based on the temperature at the current spatial location $Temp(q_j(t))$, scaled by the maximum temperature in the entire domain $T_{max}$. The two fitness components are defined as follows:

$$f_d = \sum_{i=1}^{M} d\bigg(q_{j}(t_i),S\bigg), \hspace{1cm} f_{t} = \sum_{i=1}^{M} |Temp(q_j(t_i))|
$$
Where $S$ is the location of the nearest potential spawning site.  

The choice of the temperature fitness criterion is motivated by the possibility that fish may use temperature as an environmental cue during spawning migration. Fish that follow higher temperatures are more likely to arrive in the southern spawning regions. Additionally, fish tend to avoid extreme thermal conditions (too cold) that would require excessive energy expenditure for swimming. On the other hand, the distance to spawning sites fitness criterion can serve as a proxy for unknown factors, such as magnetic fields, chemical signals, or genetic traits that influence fish to swim towards their home despite strong ocean currents \cite{hubbard2004model}.

The weighting factor $\alpha$ is designed such that fish primarily prioritize following higher temperatures when they are far from the spawning regions. As they gradually approach the spawning regions, their preference for higher temperatures decreases, leading them to adjust their swimming directions accordingly.

\subsubsection*{Selection, Crossover, and Mutation}
After each complete spawning migration simulation (once per generation), the fitness function (Equation \ref{eq:fitness_fun}) is used to evaluate each agent fish, determining their success in reaching the spawning sites. The agents with higher fitness values, indicating proximity to the spawning regions, are considered the best candidates and are selected for reproduction in the next generation. Specifically, based on fitness, the top quarter of the population is chosen as potential parents. These agents, whose ANN weights produced trajectories reaching the spawning sites, are deemed favorable parents.

To create the next generation, pairs of parents, consisting of a mate and a mum, are randomly selected from the pool of best candidates. The offspring's weights (ANN weights) are calculated by recombining the mate and mum weights using a one-point crossover. The crossover occurs at a randomly chosen breakpoint, selected from the total number of nodes in the ANN.

Mutations are randomly applied to each ANN weight with a 1\% 

This three-step process (selection, crossover, and mutation) is repeated to create several offspring equal to the initial population size, thus generating a new generation of agents.

\begin{algorithm}
\caption{Parallel Genetic algorithm}\label{alg:cap}
\begin{algorithmic}[1]
\State \textbf{Initialize ParGA(}\textit{InputParams}{)}
\While {max generations or optimal solution is found}
    \ParFor{$i \gets 1$, NumSubPopulations}
    
            \State \textbf{Evaluate} Fitness(All indiv $\in$ SubPopulations\{\textit{i}\} );
            \State \textbf{Select} BestSubPop $\gets$  \text{The best 25}$\% $ \text {of SubPopulations}\{\textit{i}\}
            
            \For{$j \gets 1$, NumSubPopulations}
                    \State \textbf{Select} randomly two different individuals (\textit{Mom,Mate}) $\in$ \textit{BestSubPop} ;
                    \State \textit{AuxIndiv} $\gets$ \textbf{Recombination}(Mom,Mate);

                    \If {\textit{rand}$\le$ \text{mutation rate}} 
                    
                            \State \textit{AuxIndiv} $\gets$ \textbf{mutation}(AuxIndiv);
    
                    \EndIf
                    \State NewSubPopulations\{\textit{j}\} $\gets$ AuxIndiv;
        \EndFor
        \State NewPop\{\textit{i}\} $\gets$ NewSubPopulations;
        \State SubPopulations\{\textit{i}\} $\gets$ NewSubPopulations;
    \EndParFor
\EndWhile
\end{algorithmic}
\end{algorithm}

\section{Simulations, validations, and sensitivity analysis }
\label{sec.4}
\subsection*{Capelin Spawning Migration}
We conducted simulations to model the migratory behavior of Barents Sea capelin during their spawning migration towards the northern Norwegian coast in 2019. The agents were initially released in the overwintering region of capelin, based on winter observational surveys \cite{Johanna2020,alrabeei2021spatial}. The simulation period spanned from January 10th to March 31st. The agents were released five times, starting from January 10th and continuing until February 13th, covering the entire historical potential migration timing \cite{gjosaeter1998population}. The release pattern followed a gradual increase in the number of released agents, reaching a peak in the third release in late January and decreasing again in the final release.

\subsection*{Model Training}
We utilized behavioral responses as a criterion to assess the convergence of the genetic algorithm (GA) during the training simulations. Convergence, in terms of behavior, was evaluated based on how well the trained model captured the aggregation of individuals in the spawning areas. Two stopping criteria were applied to terminate the GA adaptation process. The first criterion was achieving acceptable solutions, defined as 90\% of the initial population successfully reaching their spawning regions. The remaining 10\% represent individuals who may have faced predation risks or encountered deadly extreme environmental conditions. The second stopping criterion was reaching the maximum number of generations, set arbitrarily to 300. This criterion was applied when the solution no longer improved, indicated by a plateau in the ratio of successful spawners (arrival) without further increase.
\subsection*{Model Validation}
We compared our results with data from the capelin spawning migration survey to validate the predicted spawning sites. This survey involves trawl-acoustic monitoring of the spawning stock of capelin during their migration to the coast. Conducted by the Institute of Marine Research (IMR), the survey covers predefined historical spawning areas off the Troms and Finnmark coast between late February and early March \cite{skarettesting2020}. We compared the simulated fish aggregation and abundance in the spawning region with the data obtained from the observational survey.
\subsection*{Sensitivity Analysis}
Given precise knowledge of when and where the spawning fish initiate their migration, we conducted sensitivity analyses on our model, specifically concerning the migration's initial locations and starting date.

For the sensitivity analysis of initial locations, we tested two different latitudinal degrees based on the observed distribution of overwintering mature capelin stocks in the winter surveys. We simulated the migration starting from the northernmost and southernmost observed distribution points, corresponding to latitudes 74$^\circ$N and 76$^\circ$N, respectively.

Regarding the sensitivity test on starting the migration, we maintained the same initial locations (eight centers of gravity (CoGs) of the observed distribution in January and February). However, we evenly released the agent capelin multiple times during the simulation period to observe the effects of different migration initiation times.
\section{Results and Discussion}
\label{sec.5}
\subsection{Capelin Migratory Behavior}
\label{subsubsec.traj}
\begin{figure}[!ht] 
\centering
\begin{subfigure}{0.4\textwidth}
\includegraphics[scale=0.3]{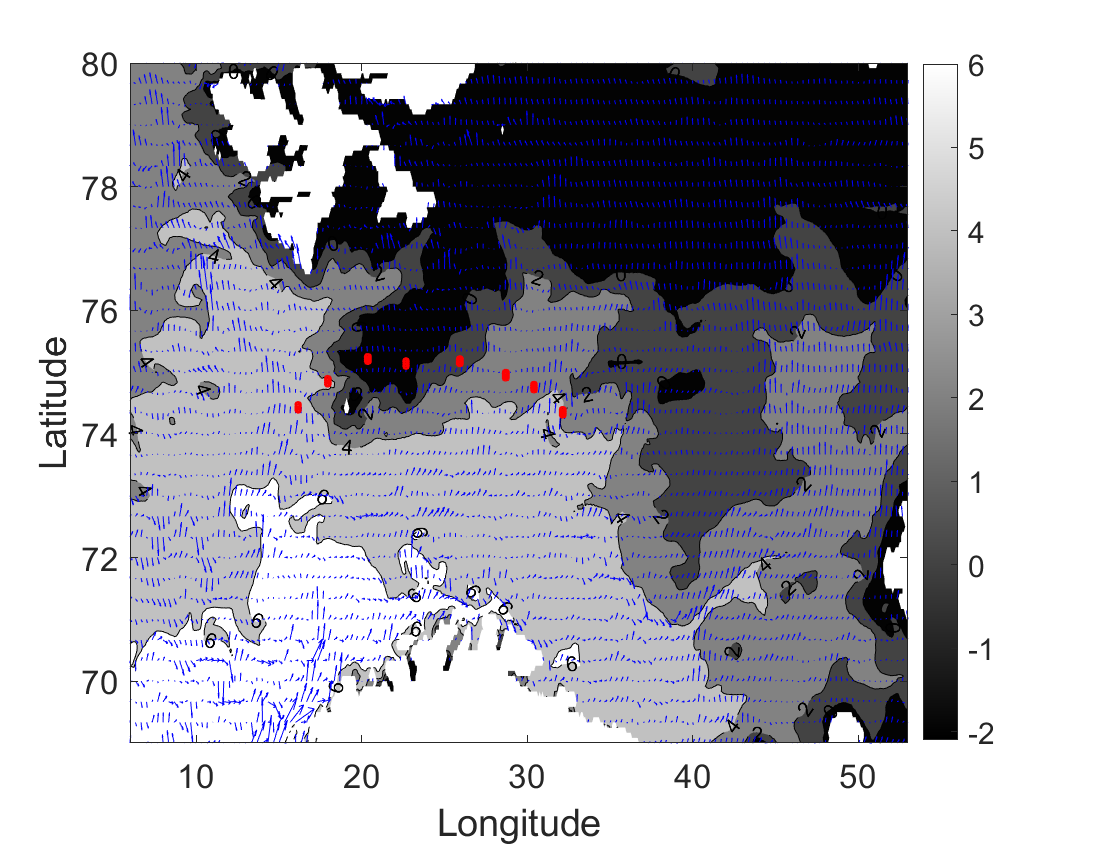} 
\caption{January  10 (first release day)} \label{fig:a1}
\end{subfigure}
\begin{subfigure}{0.4\textwidth}
\includegraphics[scale=0.3]{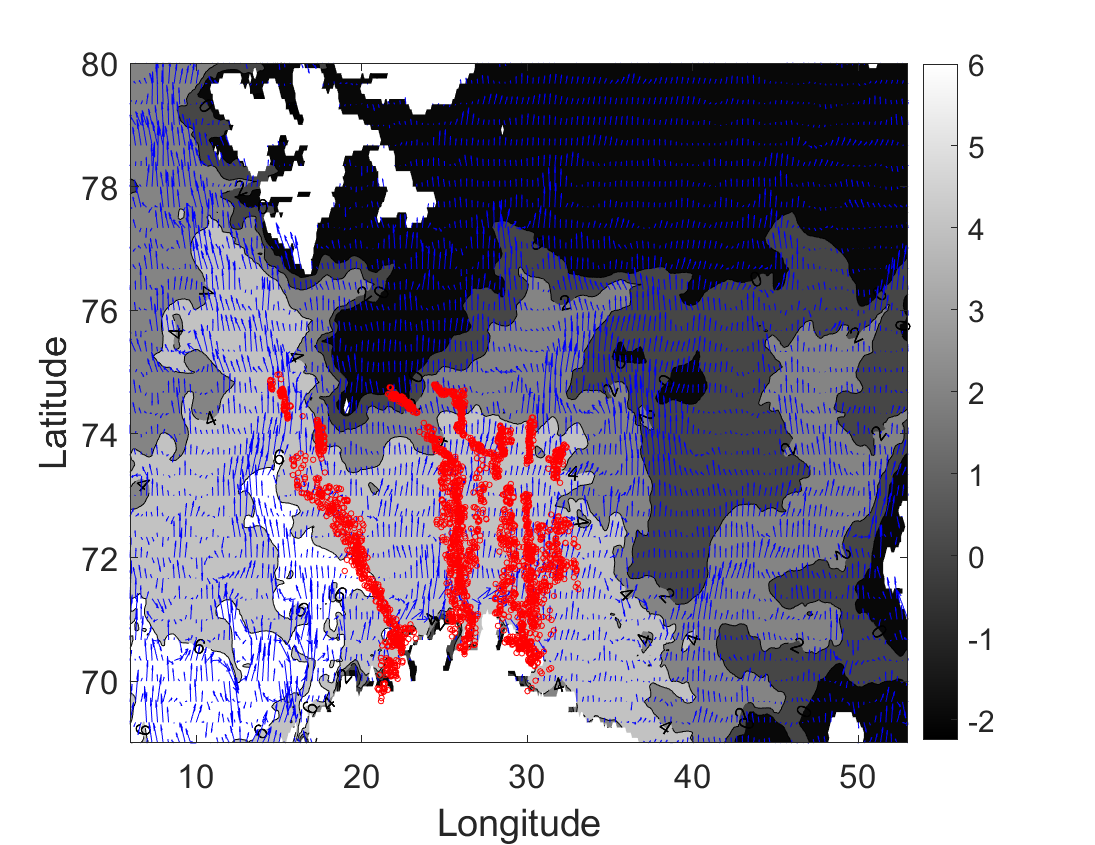} 
\caption{January 28} \label{fig:b1}
\end{subfigure}
\medskip
\begin{subfigure}{0.4\textwidth}
\includegraphics[scale=0.3]{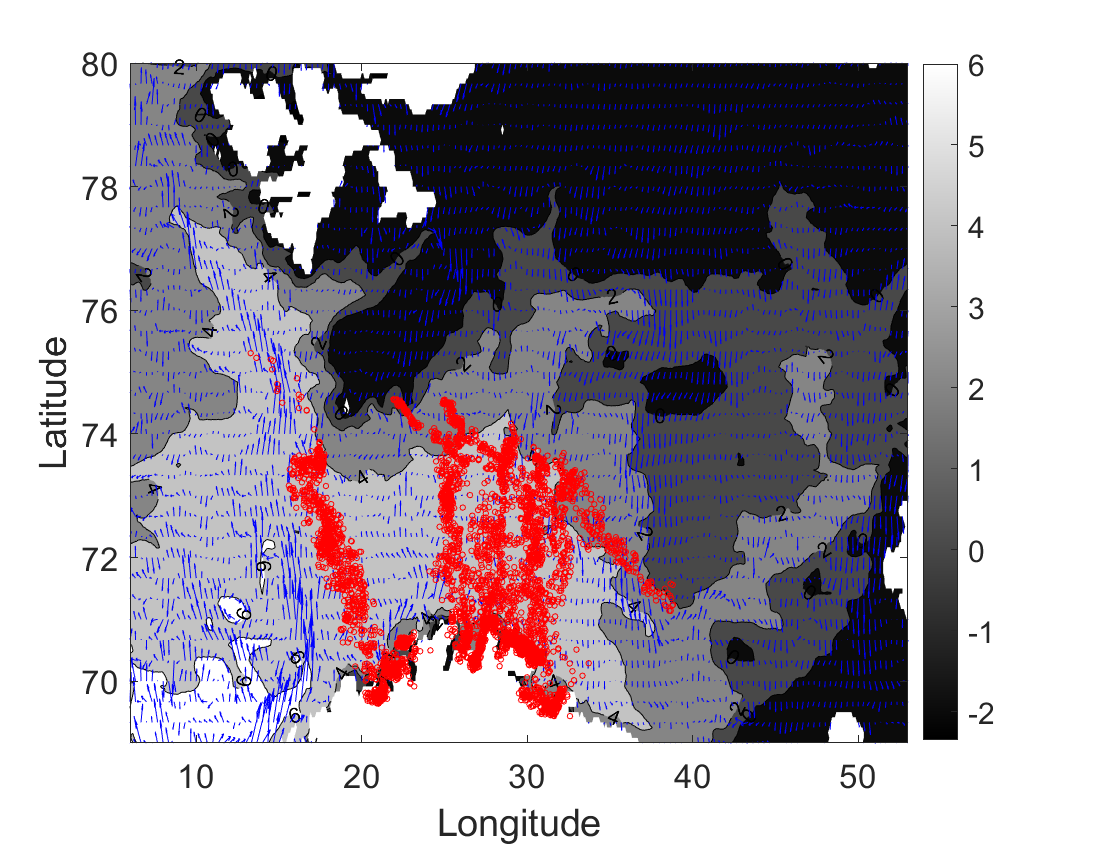}
\caption{February 17}  \label{fig:c1}
\end{subfigure}
\begin{subfigure}{0.4\textwidth}
\includegraphics[scale=0.3]{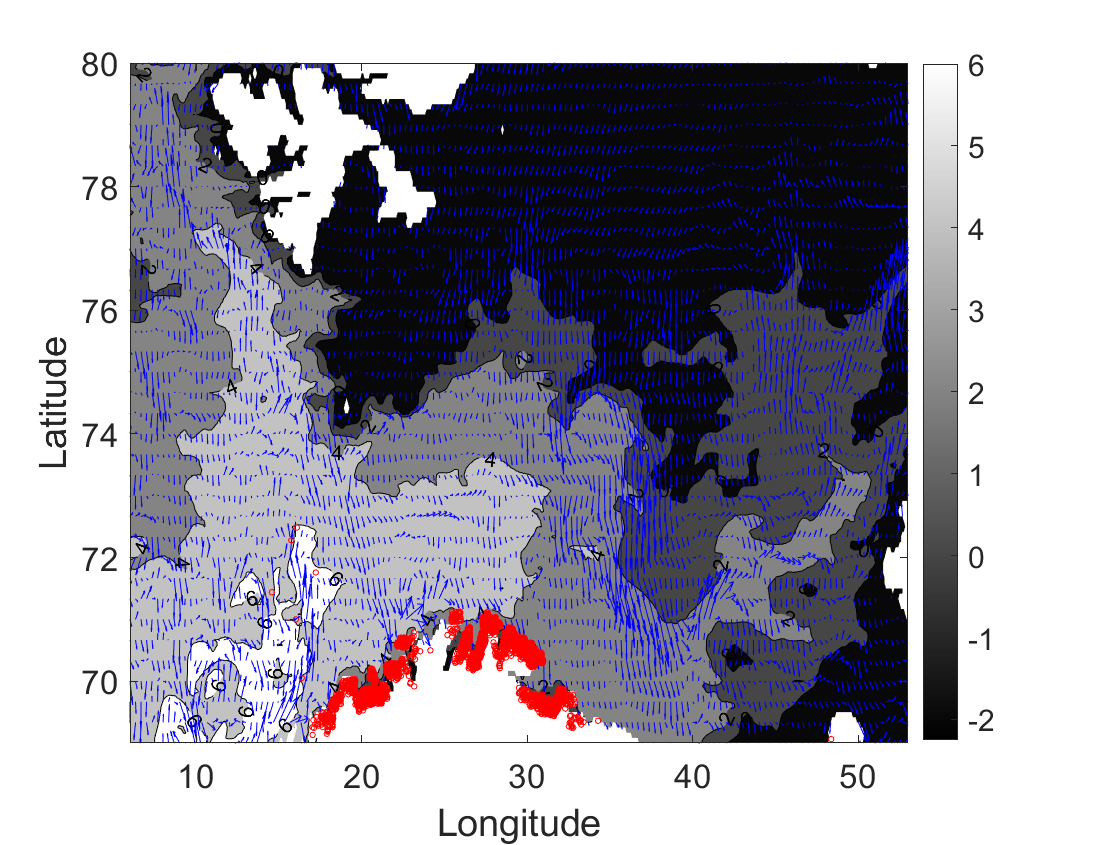}
\caption{March 31} \label{fig:d1}
\end{subfigure}
\caption{The distribution of the simulated capelin is indicated by the red dots, while the blue arrows represent the daily mean near-surface current velocity. The contour lines represent the daily near-surface water temperature, with colors ranging from black at low temperatures to white at high temperatures.} \label{fig:Active_swimming1}
\end{figure}

Using the adaptive agent model, we successfully simulated the migratory behavior of capelin toward their spawning grounds. The model accurately captured the southeastward movement of capelin during their spawning migration (Figures \ref{fig:a1} to \ref{fig:c1}). The majority of spawning activity occurred over the historical spawning sites located along the eastern coast of northern Norway, east of $20^\circ$E (Figure \ref{fig:c1}). The adaptive artificial neural networks (ANNs) effectively replicated the observed distribution of spawning fish over the spawning grounds, aligning well with the data from the capelin spawning migration survey conducted by Skaret et al. \cite{skarettesting2020}.

More generally, this result indicates that the combination of artificial neural networks (ANNs) trained by a genetic algorithm (GA) with a well-defined fitness function holds promise for simulating fish migrations in scenarios where multiple variables influence migratory behaviors or when the dominant environmental driver of migration is not clearly understood. By optimizing the weights of the ANNs through the GA, the model was able to capture the complex dynamics of capelin migration, taking into account factors such as environmental conditions and distance to spawning sites. This demonstrates the potential of using ANNs and GA-based approaches to study and predict migratory behaviors in fish populations.
 \begin{figure}[!ht]
    \centering
     \begin{tabular}{ll}
    \includegraphics[scale=0.3]{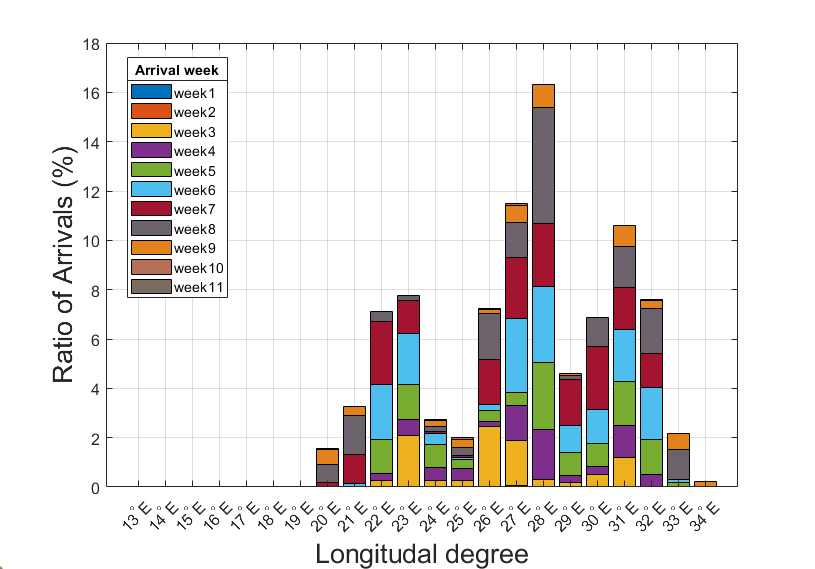}       & \hspace{-1cm}\includegraphics[scale=0.4]{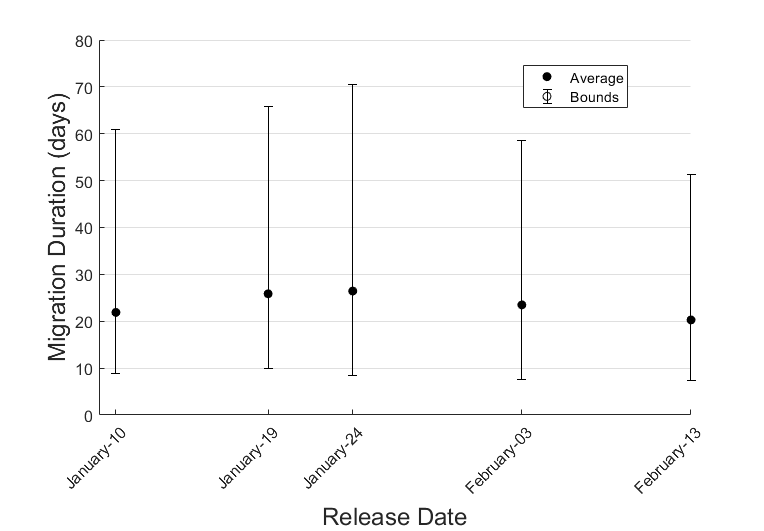}  \\
    \hspace{3cm}a.  Ratio of arrival fish      & \hspace{1cm}  b.  Migration duration (days) 
     \end{tabular} 
\caption{Summary statistics:  Left: ratio of arrivals at spawning sites located within historical areas, and Right: statistics of simulated migration duration for different release dates.} \label{fig:ratio_week_location}
\end{figure}
%
\subsection{Models comparison}
\subsubsection*{\bf Experiment-1: Passive swimmers}
\label{subsubsec.exper1}
\begin{figure}[ht!] 
\begin{subfigure}{0.48\textwidth}
\centering
\includegraphics[scale=0.3]{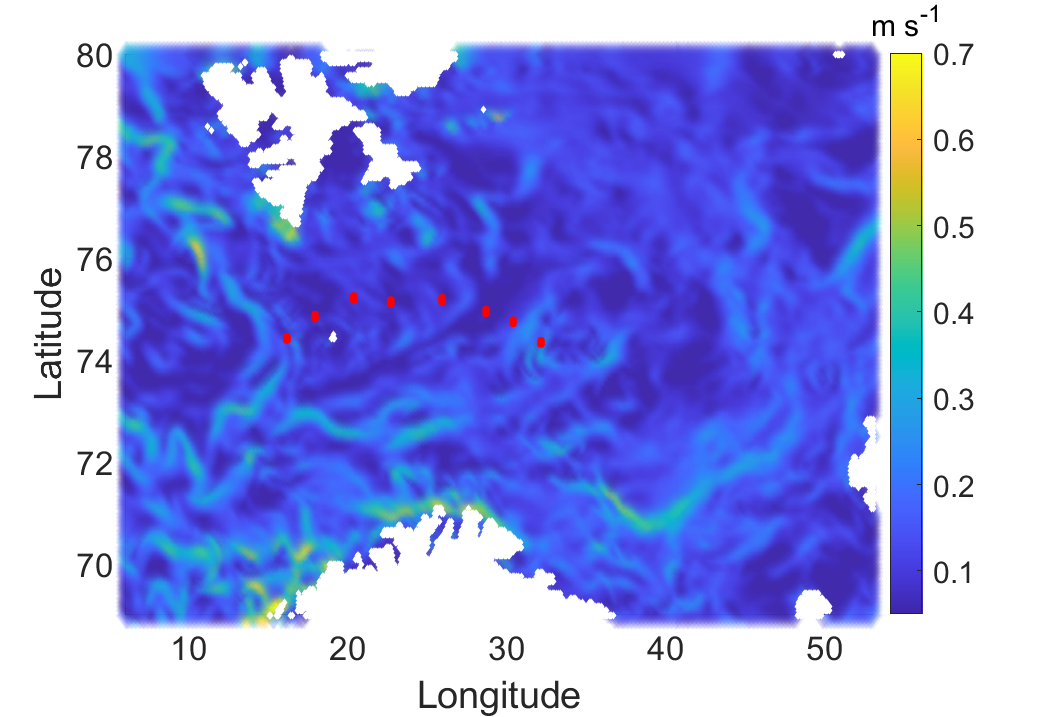} 
\caption{January  9 (first release day)} \label{fig:ac}
\end{subfigure}
\begin{subfigure}{0.48\textwidth}
\centering
\includegraphics[scale=0.3]{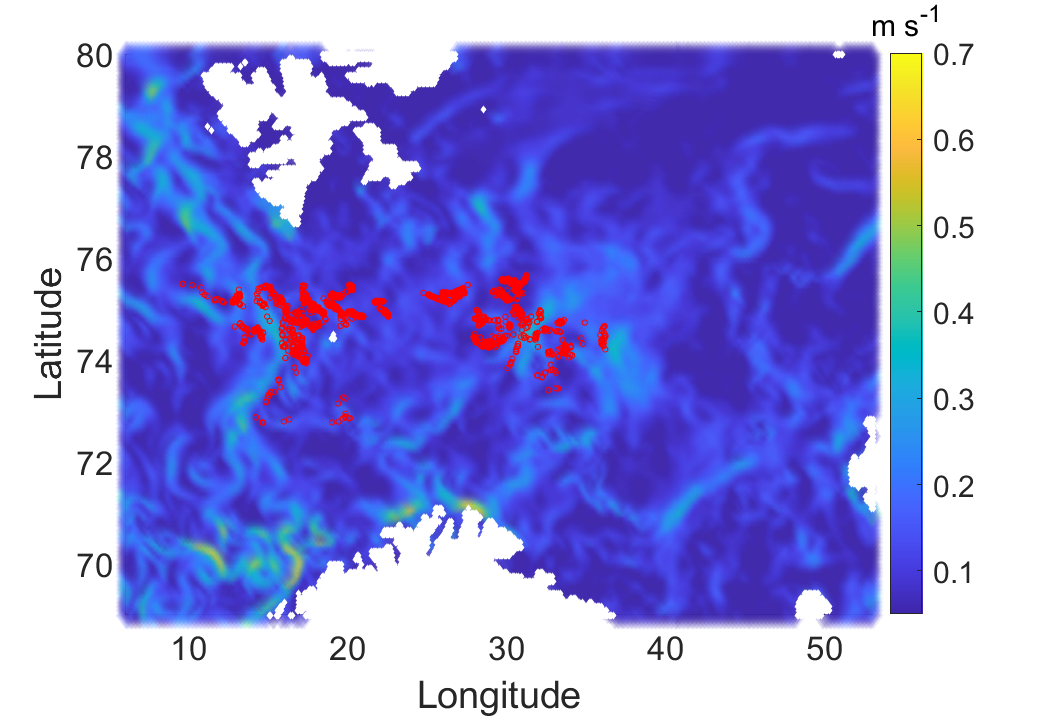}
\caption{January 28} \label{fig:bc}
\end{subfigure}
\medskip
\begin{subfigure}{0.48\textwidth}
\centering
\includegraphics[scale=0.3]{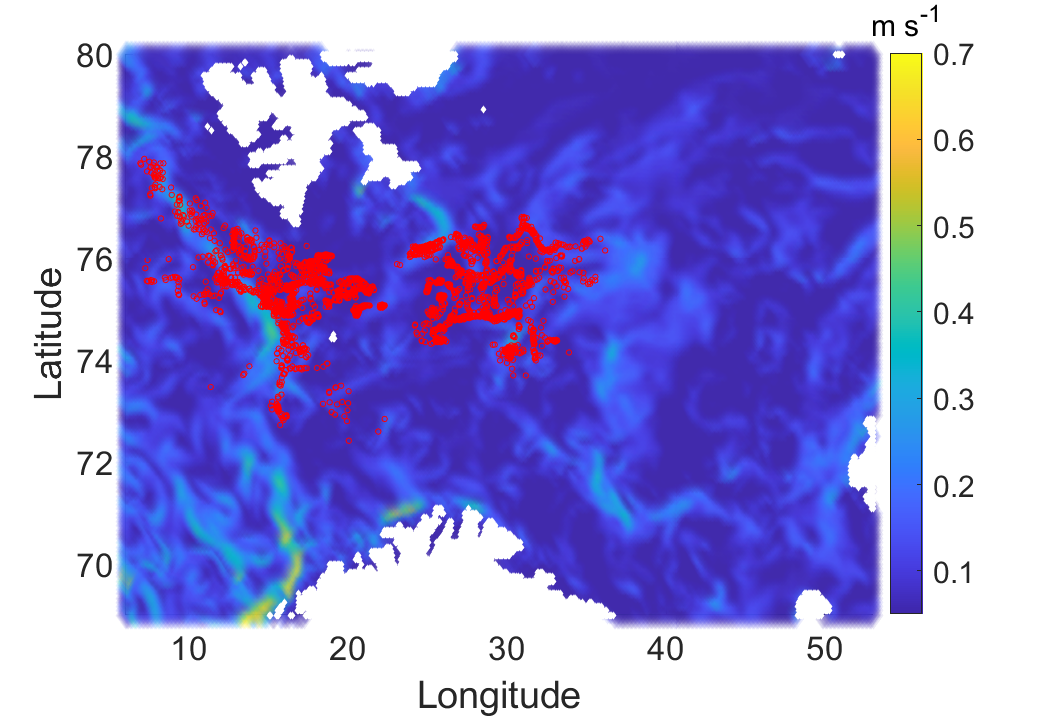}
\caption{February 17} \label{fig:cc}
\end{subfigure}
\begin{subfigure}{0.48\textwidth}
\centering
\includegraphics[scale=0.3]{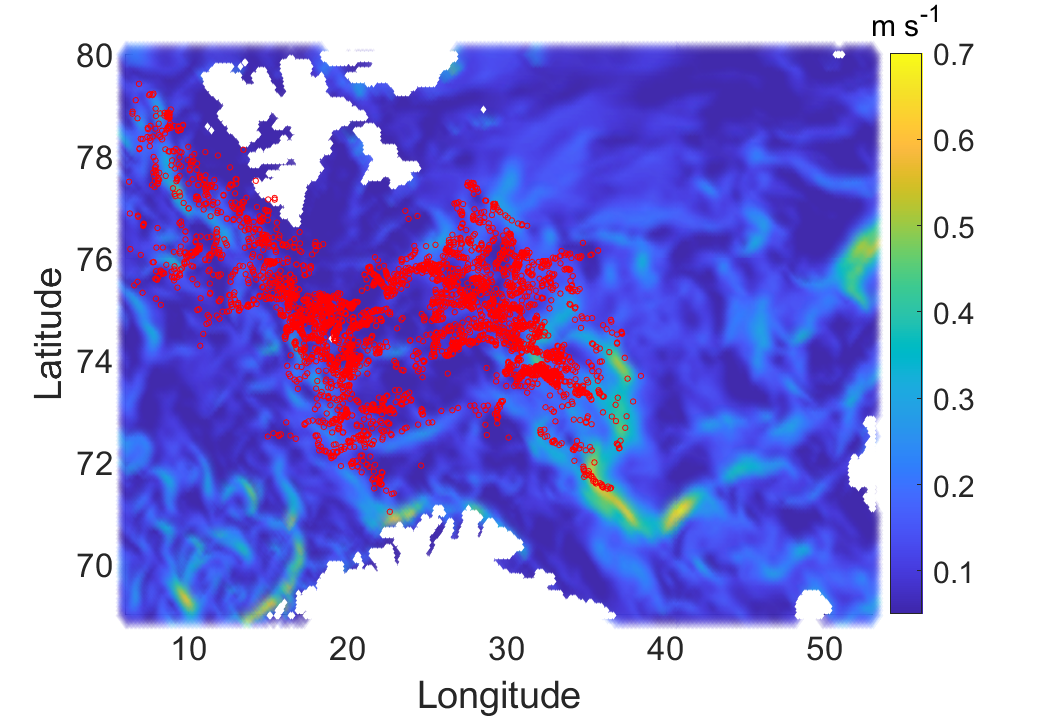}
\caption{March 31} \label{fig:dc}
\end{subfigure}
\caption{The spatial distribution of capelin when they are passively advected by the Barents Sea currents, without active swimming, during the spawning migration. The red dots represent the simulated fish, while the background depicts the daily scalar absolute value (magnitude) of Barents Sea averaged current velocity at the first 50 meters depth.
} \label{fig:current_advection}
\end{figure}

In order to investigate whether capelin fish can be considered passive swimmers, we conducted an experiment where the decision-making process was removed from the simulation presented in Section \ref{subsubsec.traj}. In this experiment, the direction of the agents was solely determined by the dispersion of ocean current velocities ($Vx_{k}=Vy_{k}=0$). The result of this experiment is presented in Figure (\ref{fig:current_advection}) 
The results of this experiment, shown in Figure \ref{fig:current_advection}, highlight the significant influence of ocean currents on capelin spawning migration trajectories. In the western part of the Barents Sea, where the Atlantic current prevails \cite{ingvaldsen2013responses}, a large proportion of the simulated fish were observed to drift northwestward along the coast of Svalbard. Only a small fraction of the fish were transported southward toward the spawning sites in the Finnmark region. On the other hand, fish released from the eastern part of the Barents Sea were observed to be drifted east-northward by the Atlantic current.

Furthermore, it was observed that the final locations of the fish in this experiment were unable to reach latitudes above approximately $71^{\circ}$N between longitudes $28-35^{\circ}$E, corresponding to the coast of Finnmark and the Kola Peninsula. Although a small number of fish were able to reach latitudes up to $71^{\circ}$N in the western part of longitude $28^{\circ}$E, this represented only a minor fraction of the overall migrating fish. These findings lead to the conclusion that the spawning migration of capelin is not solely dependent on passive drifting, but rather assisted by the ocean currents, indicating an active role of the fish in their migratory behavior.

\subsubsection*{\bf Experiment-2: Taxis movement}
\label{subsubsec.exper2}
\begin{figure}[htb!] 
\begin{subfigure}{0.31\textwidth}
\centering
\includegraphics[scale=0.32]{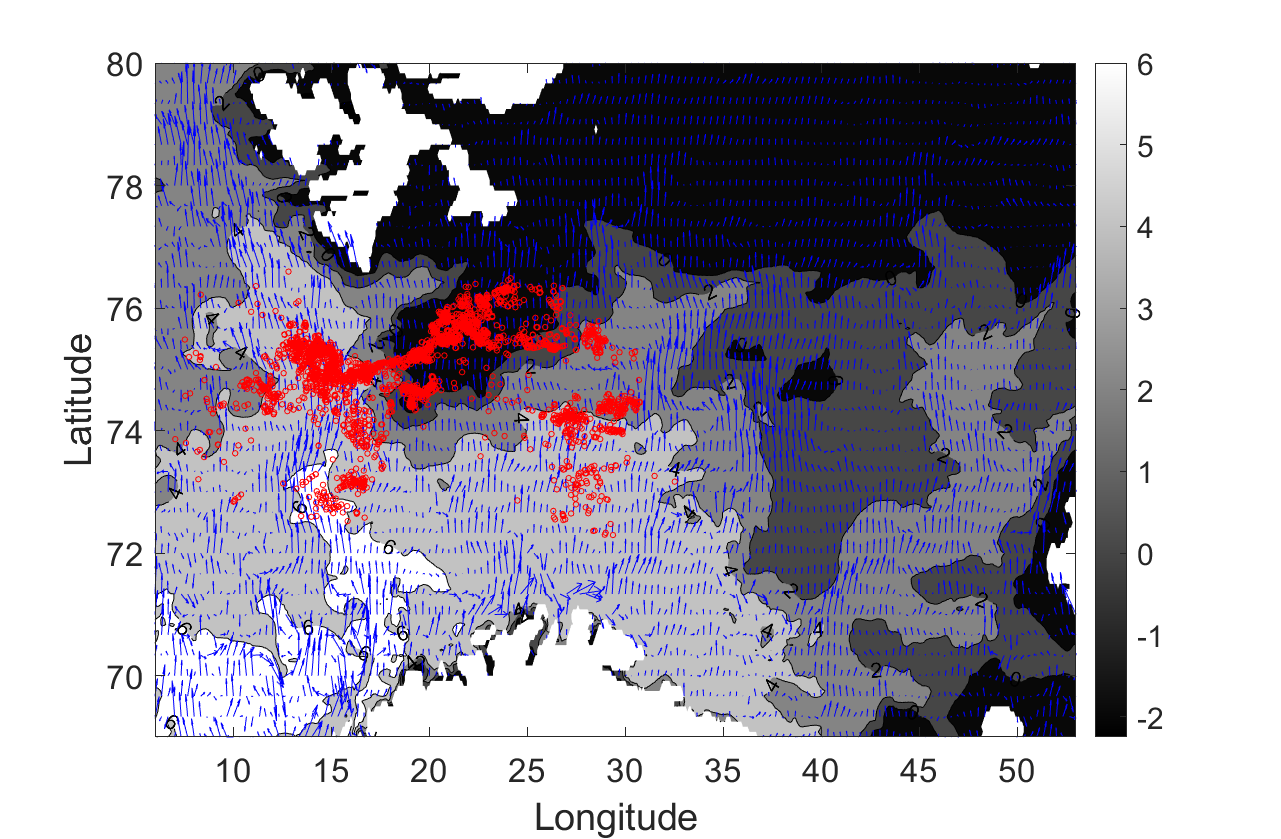} 
\caption{January 28} \label{fig:ag}
\end{subfigure}
\begin{subfigure}{0.31\textwidth}
\centering
\includegraphics[scale=0.32]{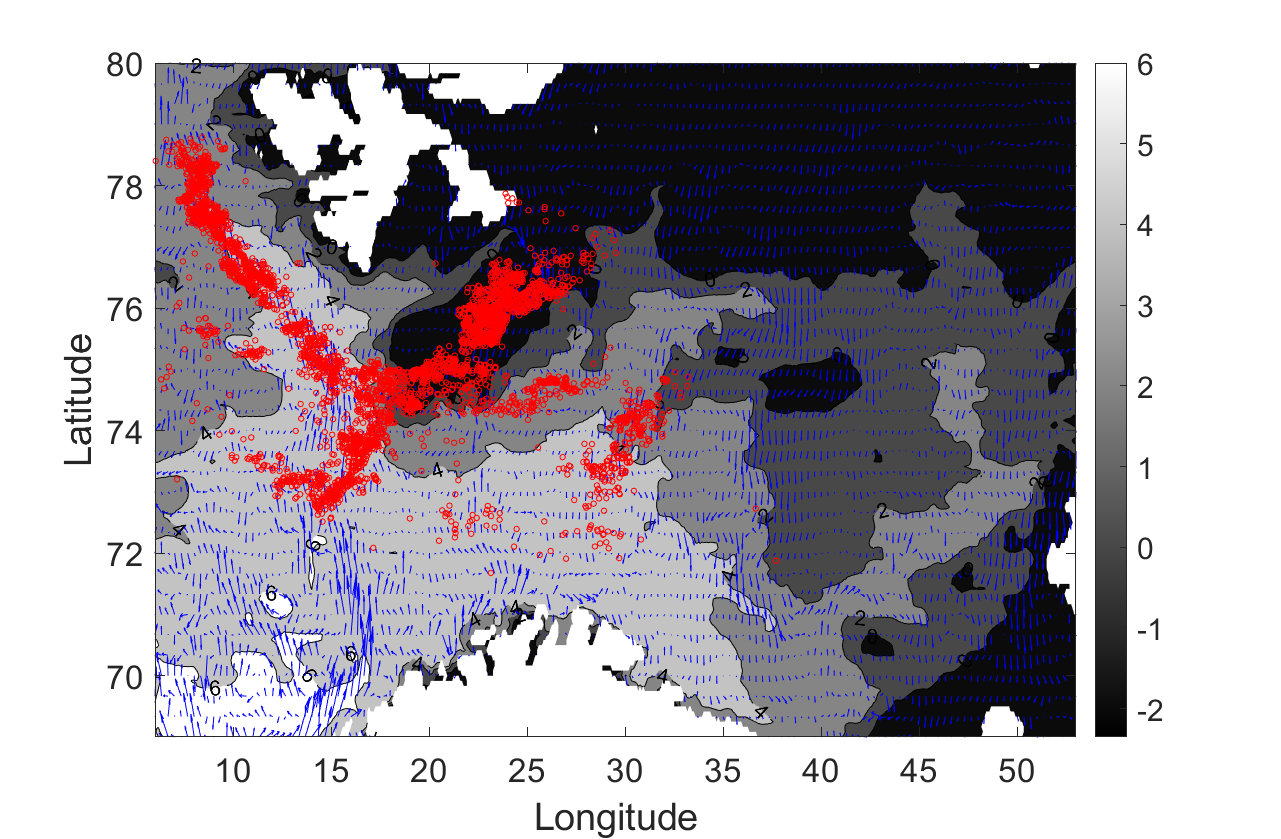} 
\caption{February 17} \label{fig:bg}
\end{subfigure}
\begin{subfigure}{0.31\textwidth}
\centering
\includegraphics[scale=0.32]{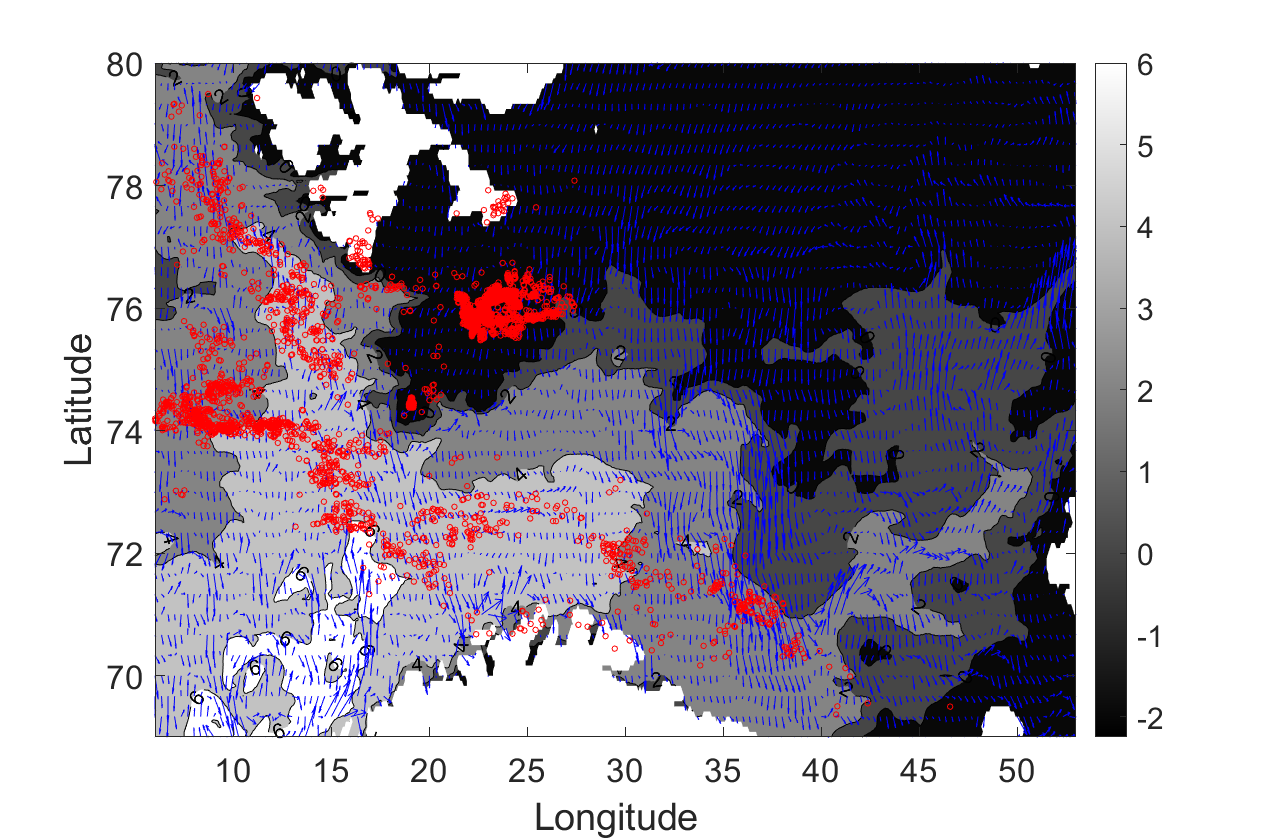}
\caption{March 31} \label{fig:cg}
\end{subfigure}
\caption{The distribution of the simulated capelin, represented by red dots, is depicted while utilizing the temperature gradient detection model (taxis model). The blue arrows denote the daily mean near-surface current velocity, and the contour illustrates the daily near-surface water temperature. The color scale ranges from black at lower temperatures to white at higher temperatures.} \label{fig:Active_swimming}
\end{figure}

To investigate whether sea surface temperature alone is sufficient to control capelin movement toward its spawning grounds, we implemented a temperature gradient detection model in this experiment. We utilized two well-known models that have been previously used to simulate fish migration. It is important to note that this approach does not involve learning by the ANN or adaptation by the GA algorithms. In this experiment, the capelin agents determined their next directional movement based solely on the temperature gradient of the surrounding environment(\ref{fig:ag}-\ref{fig:cg}).

The results of this experiment, presented in Figures \ref{fig:ag} through \ref{fig:cg}, demonstrate the distribution of the simulated capelin using the temperature gradient detection model (taxis model). Blue arrows represent the daily mean near-surface current velocity, while the contour depicts the daily near-surface water temperature. The color range varies from black at low temperatures to white at high temperatures.

While the simulated fish using the gradient detection model were observed to move towards the spawning region at a faster pace compared to the results in Experiment 1 (passive swimming), the majority of the fish failed to successfully reach the spawning regions by the end of the simulations (Figure \ref{fig:cg}). A significant proportion of the fish drifted northwestward, carried by the strong current in the western part of the Barents Sea (around 17$^{\circ}$E). This happened despite the SST gradient, which indicated that movement towards the south, where it is warmer, should be preferred. This finding suggests that the temperature gradient detection model may not accurately represent the spawning migratory behavior of capelin.
\subsubsection*{\bf Experiment-3: Restricted-area search}
\label{subsubsec.exper3}
The purpose of this experiment was to investigate the influence of the amount of information available to the capelin agents on their behavior. In this experiment, we kept the same stimulus (temperature) as described under Experiment 2 (taxis movement) but introduced a scenario where the agents possessed knowledge of their surroundings and would adjust their movements towards the location with the highest temperature. Consequently, we re-executed the simulation described in Section \ref{subsubsec.exper2}, but we made adjustments to the agent model to have them select the cell with the highest temperature from among the eight nearest cells in the mesh.

\begin{figure}[H] 
\begin{subfigure}{0.33\textwidth}
\includegraphics[scale=0.25]{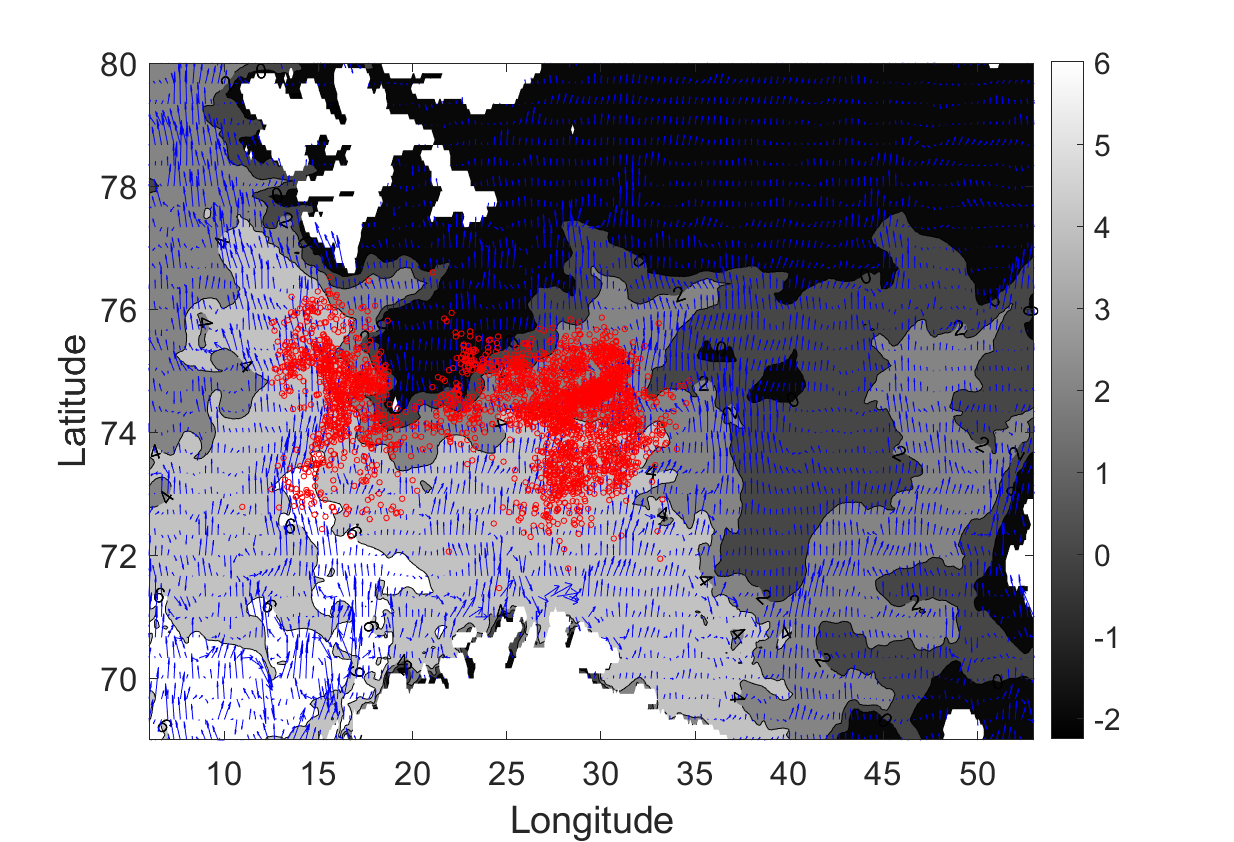} 
\caption{January 28} \label{fig:dr}
\end{subfigure}
\begin{subfigure}{0.33\textwidth}
\includegraphics[scale=0.25]{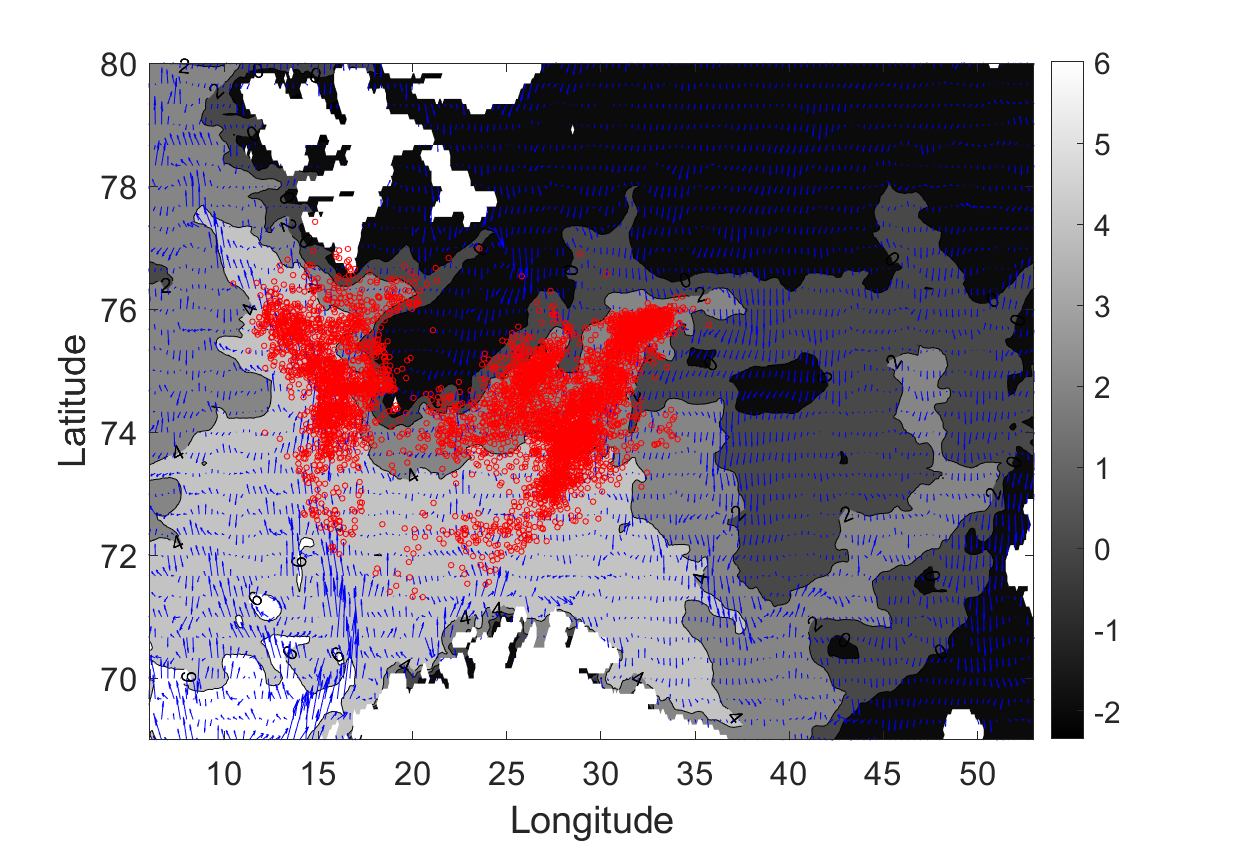} 
\caption{February 17} \label{fig:er}
\end{subfigure}
\begin{subfigure}{0.33\textwidth}
\includegraphics[scale=0.25]{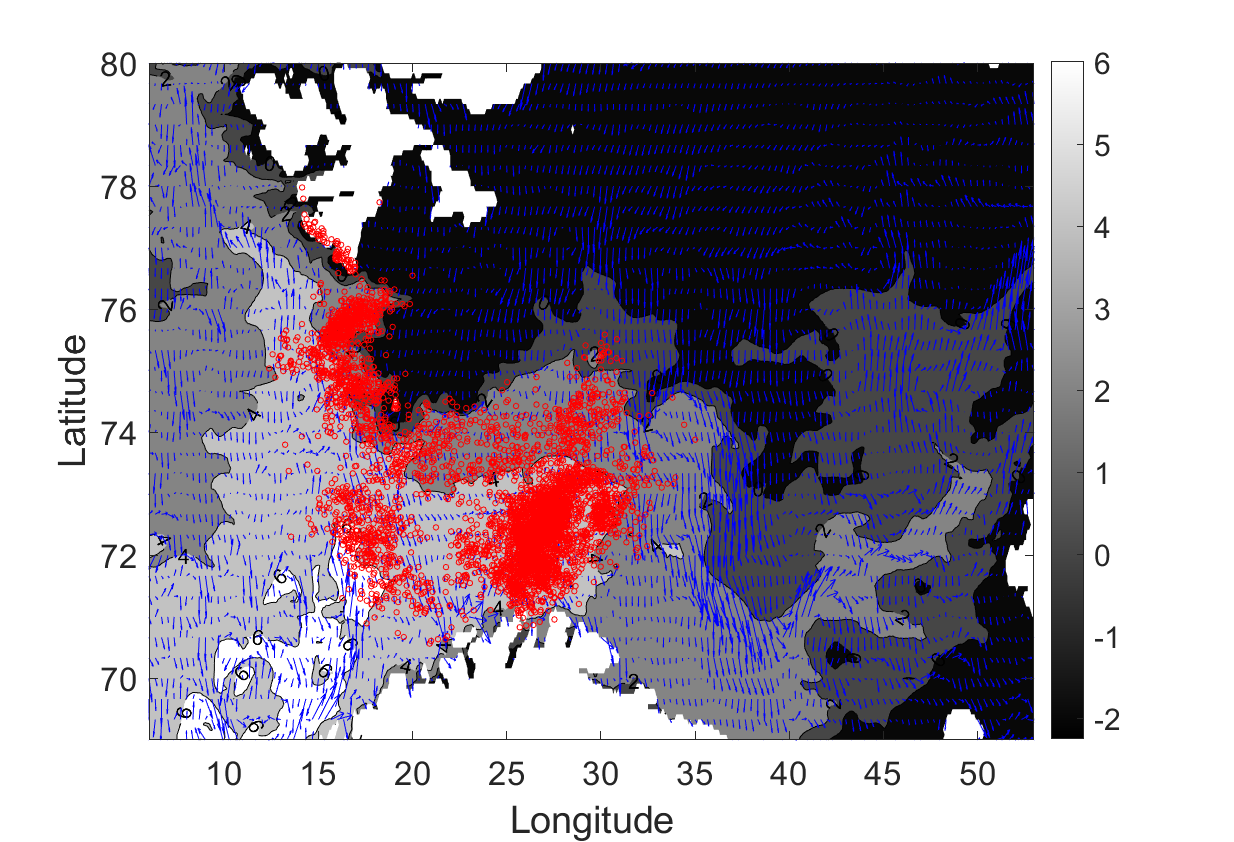}
\caption{March 31} \label{fig:fr}
\end{subfigure}
\caption{The distribution of the simulated capelin, represented by red dots, is shown in three sets (a-c) corresponding to the temperature gradient detection model, another three sets (d-f) for the local-area search model, and the final three sets (g-i) for the optimized neural network model. The blue arrows indicate the daily mean near-surface current velocity, while the contour depicts the daily near-surface water temperature, with colors ranging from black at lower temperatures to white at higher temperatures.} \label{fig:Active_swimming3}
\end{figure}

The results of this simulation are presented in Figures \ref{fig:dr} through \ref{fig:fr}. In contrast to the gradient detection model, the simulated fish displayed a tendency to move southward and southwestward in the local search model. However, despite a small portion of the migrating population successfully reaching the spawning region, the majority of migrating fish were unable to reach the spawning region within the simulation period. In this model, larger temperature differences between cells could potentially result in more directional movement. Nevertheless, due to the limited spatial variations in sea surface temperature within local areas, most fish became trapped in local maxima and were unable to progress further.
\subsection*{Contrasting the migration models}
\begin{figure}[H]
    \centering
    \includegraphics[scale=0.45]{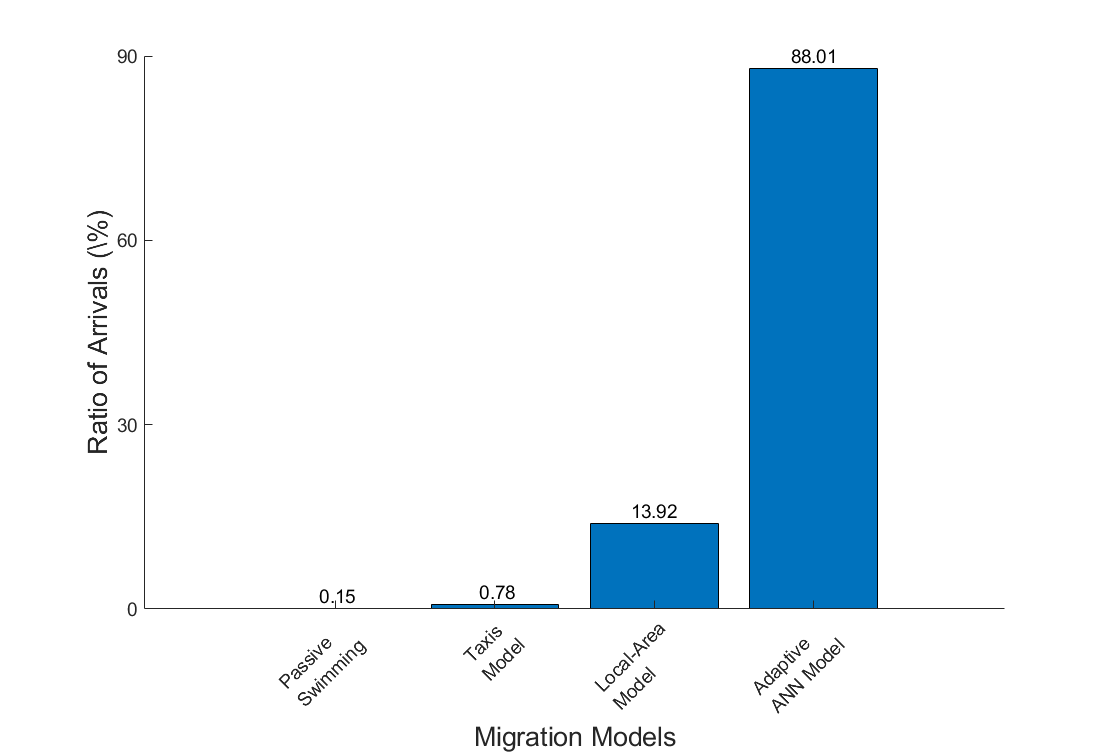} 
\caption{Summary statistics of efficiency (ratio of arrivals) using different fish migration models.} \label{fig:ratio_arrivals}
\end{figure}

To quantitatively compare the efficiency of the three modeling methods with our adaptive model, we calculated the ratio of arrivals to the spawning regions for each method. The results are presented in Figure \ref{fig:ratio_arrivals}.

In general, the simulations with active swimming behavior (our adaptive model) yielded significantly better results compared to those without active swimming behavior (Experiment-1). For instance, when the swimming behavior was active, the distribution of the simulated fish extended up to a latitude of 73$^\circ$N (as seen in Figures \ref{fig:bg} and \ref{fig:er}), which occurred approximately three weeks earlier than in simulations without swimming behavior (Figure \ref{fig:bc}).

While sea surface temperature (SST) may play a crucial role in capelin fish spawning migration, relying solely on temperature-based movement mechanisms presents challenges. As indicated in the results, following the temperature gradient approach was not the most suitable method for describing capelin spawning migration in the Barents Sea, despite its success in modeling the spawning migrations of other species, such as capelin in the Icelandic Sea \cite{hubbard2004model,barbaro2009modelling} and the Japanese anchovy in the China Sea \cite{tu2012using}. This disparity may be attributed to the specific oceanographic conditions within the capelin migration range. In the Barents Sea, capelin swims from the middle of the sea toward the coast and encounters strong currents opposing their intended destination. Therefore, capelin requires robust movement strategies to overcome these opposing currents, which cannot be accomplished solely by following temperature gradients.

The restricted-area search approach, which has proven successful in several fish migration models, also failed to accurately simulate capelin spawning migration. Despite a relatively extended simulation duration of around three months, fewer than 30\% of the total released particles (simulated fish) managed to cross 72$^\circ$N. These particles became entrapped in local temperature maxima across various regions of the domain, with only a few reaching the global maximum in the southwestern part of the domain (between 10-18$^\circ$E and 68-71$^\circ$N). Approximately 14\% of the particles reached the spawning region.

Conversely, our adaptive model leveraged multiple variables (ANN inputs) from the surrounding environment to inform migration trajectories. The genetic algorithm (GA) continuously adjusted the artificial neural network (ANN) weights to enhance the agent's movement decisions during training. Consequently, the trained ANN was capable of generating migration trajectories that maximized the GA fitness function, which, in our case, aimed to maximize the number of arrivals to the spawning region through the utilization of temperature information.

In summary, our adaptive model consistently demonstrated superior performance in simulating capelin spawning migration when compared to the other methods tested, underscoring the significance of integrating multiple variables and adaptive learning to accurately represent complex migratory behaviors.
\subsection{Sensitivity analyses}
\begin{figure}[!ht] 
\begin{subfigure}{0.5\textwidth}
\includegraphics[scale=0.375]{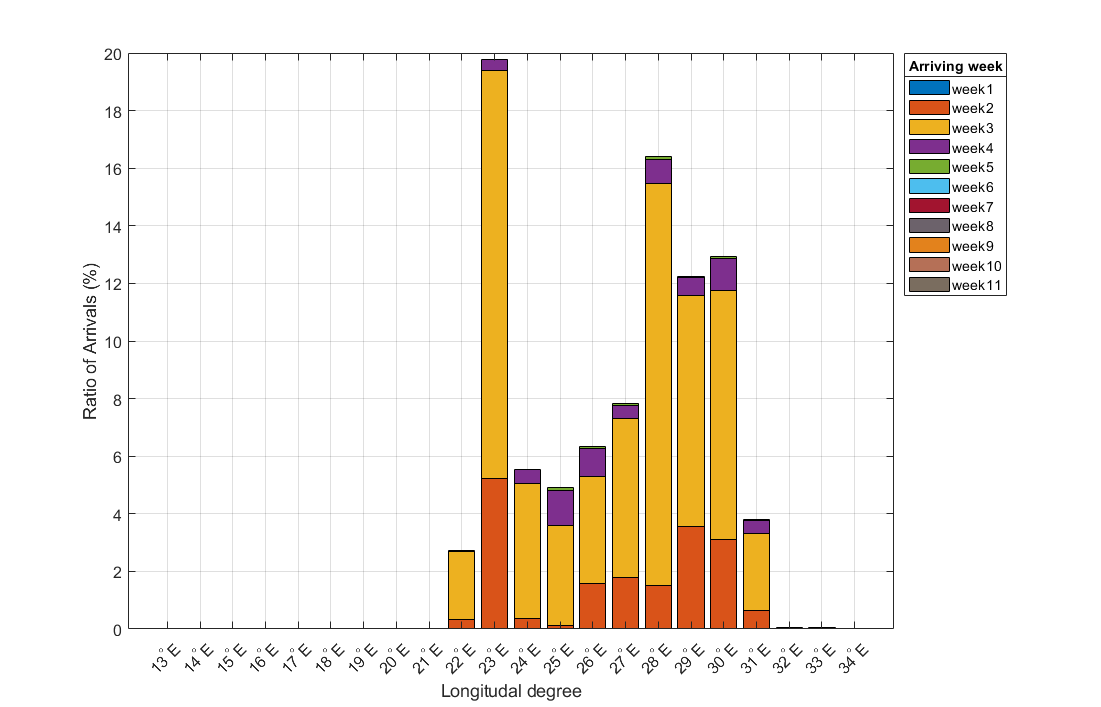} 
\caption{ 
The overall ratio of arrival is about $95\%$ (above 74$^\circ$N).} \label{fig:sin_a}
\end{subfigure}\hspace*{\fill}
\begin{subfigure}{0.5\textwidth}
\includegraphics[scale=0.375]{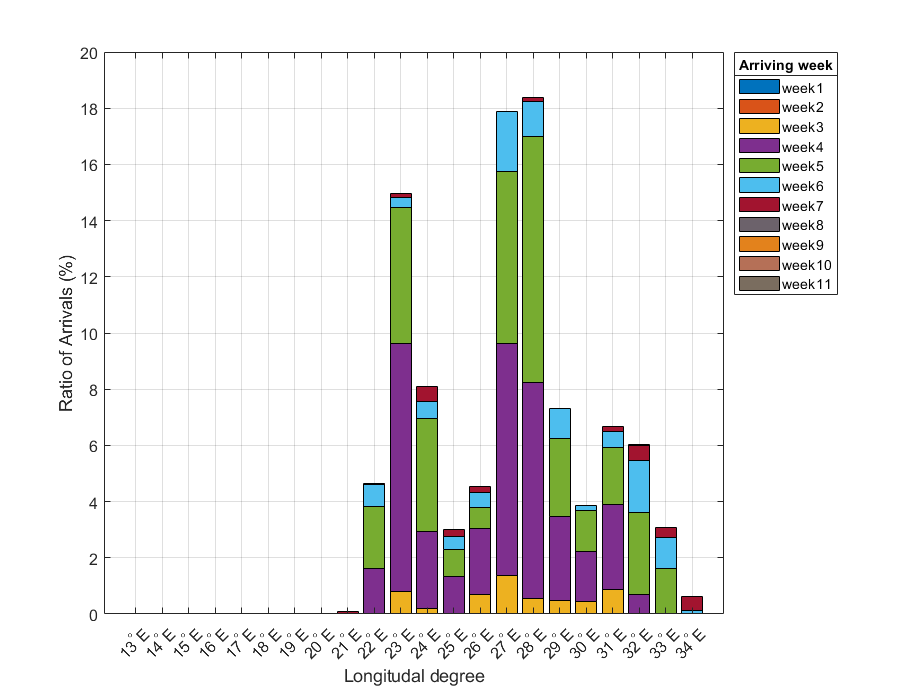}
\caption{The overall ratio of arrival is about $92\%$ (above 76$^\circ$N).} \label{fig:sin_b}
\end{subfigure}
\caption{Initial locations in the northernmost of the overwintering distribution, and ratio and distribution of the arrivals over the spawning regions. The initial locations were set along the longitudinal degrees 15-32$^\circ$E.}
\label{fig:sens_initial}
\end{figure}

The sensitivity analyses conducted in this study offer valuable insights into the various factors influencing simulated capelin spawning migration. The sensitivity analysis regarding the initial distributions reveals that capelin migration routes remain rather consistent irrespective of the initial distributions. While the abundance over the spawning sites may vary, a significant majority of simulated fish successfully reach the spawning regions. The overall ratio of arrivals was notably high, with approximately 96\% and 92\% in simulations with initial locations along the  74$^\circ$N and 76$^\circ$N degrees, respectively (see Figure~\ref{fig:sens_initial}). The coastal current may contribute to variations in distribution over the spawning zone. The difference in the ratio of arrivals between the two sets of initial locations could be attributed to fish migrating from 76$^\circ$N experiencing stronger drifting currents compared to those migrating from 74$^\circ$N.

\begin{figure}[ht!] 
\begin{subfigure}{0.5\textwidth}
\includegraphics[width=\linewidth]{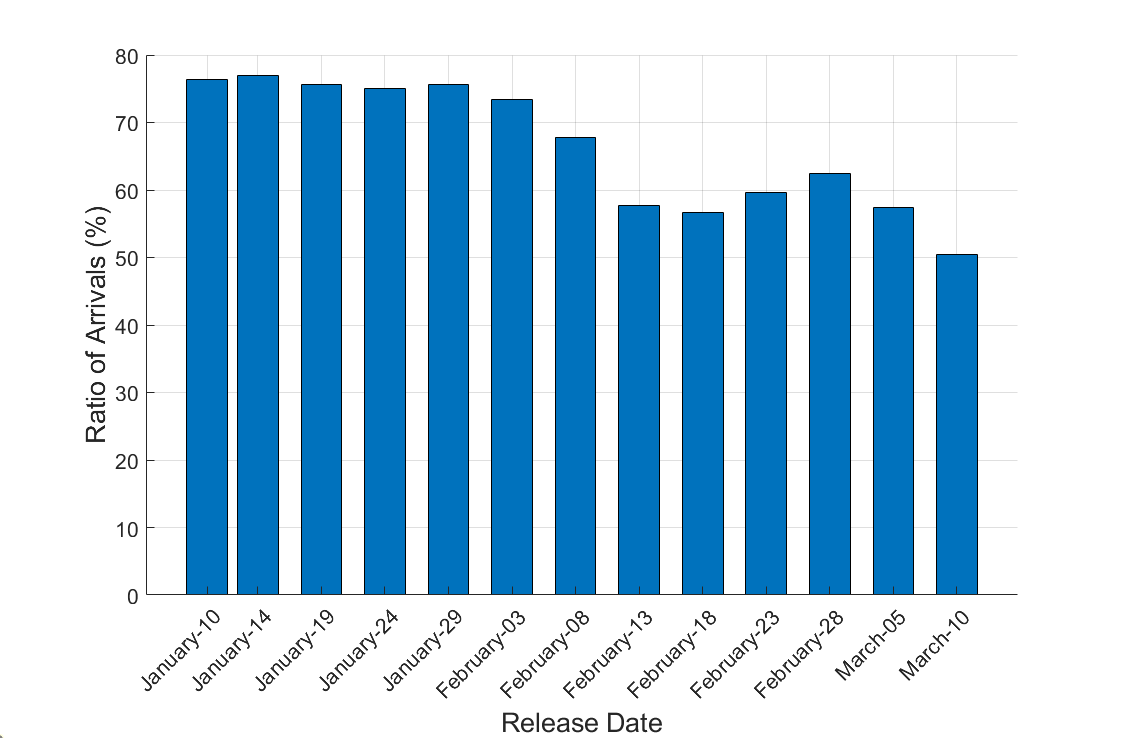} 
\caption{ratio of arrivals based on different release dates.} \label{fig:sens_arrivals_a}
\end{subfigure}\hspace*{\fill}
\begin{subfigure}{0.5\textwidth}\includegraphics[width=\linewidth]{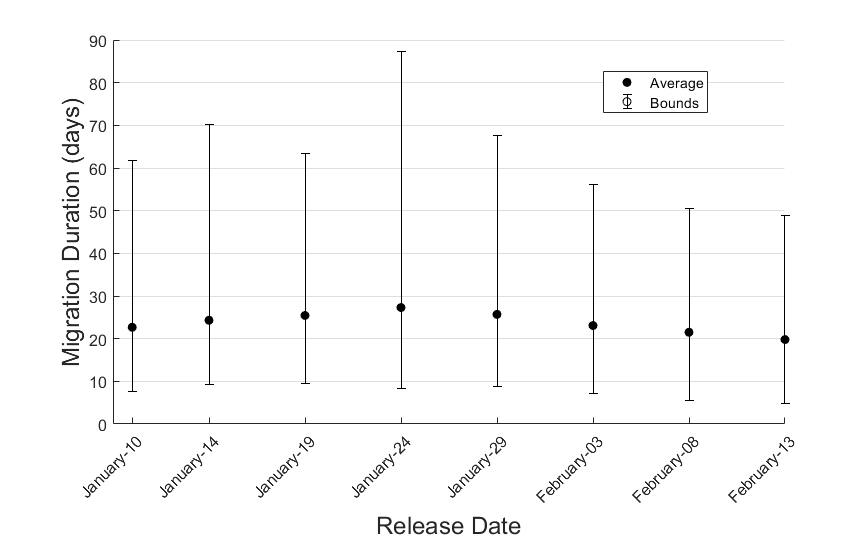}
\caption{Migration duration based on different release dates.} \label{fig:sens_arrivals_b}
\end{subfigure}
\caption{Summary statistics of simulations with different release dates.}
\label{fig:sens_timing}
\end{figure}

The sensitivity analysis concerning the releasing date (initiating the simulation) indicates that the ratio of arrivals is highest (around 76\%) when the migration begins in mid-January. However, the ratio remains substantial (approximately 75-67\%) when the migration starts in mid-January or early February. The ratio decreases by almost half when the migration is delayed until mid-February.  In contrast, the migration duration remains fairly consistent, lasting around 22-25 days, regardless of the release date (as shown in Figure \ref{fig:sens_arrivals_b}).

In terms of sensitivity to the fitness function, the choice of the fitness function in the evolutionary algorithm is a critical and challenging aspect when applied in behavioral ecology, as noted in previous research \cite{calvez2005automatic}. In our study, the fitness function played a pivotal role in training the artificial neural network (ANN) to generate realistic spawning migration trajectories. To highlight the significance of selecting an appropriate fitness function, we compared the performance of the genetic algorithm (GA) using our adaptive fitness function (Equation \ref{eq:fitness_fun}) with another test function.

\begin{figure}[H] 
\begin{subfigure}{0.33\textwidth}
\includegraphics[scale=0.375]{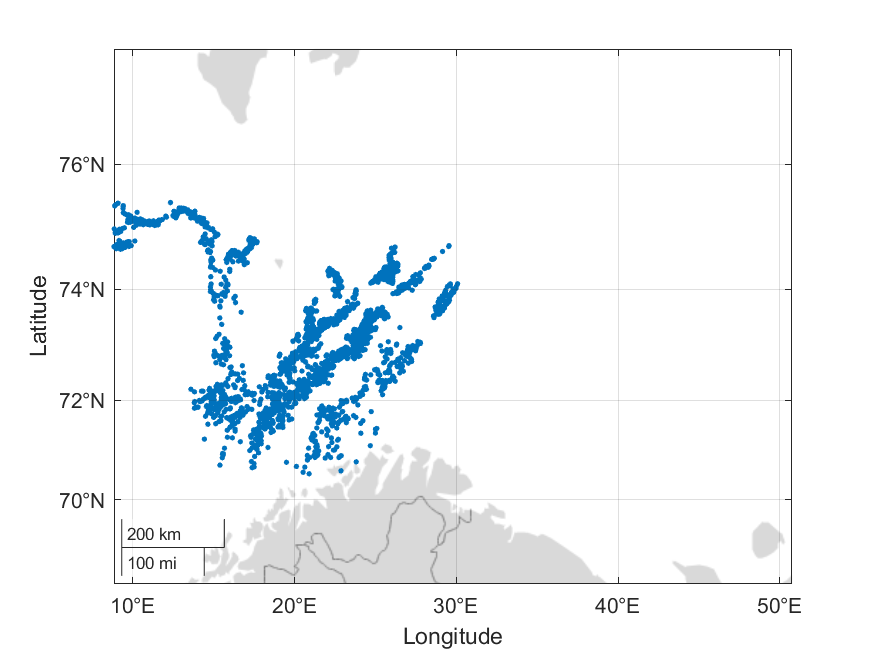} 
\caption{January 28} \label{fig:d2}
\end{subfigure}
\begin{subfigure}{0.33\textwidth}
\includegraphics[scale=0.375]{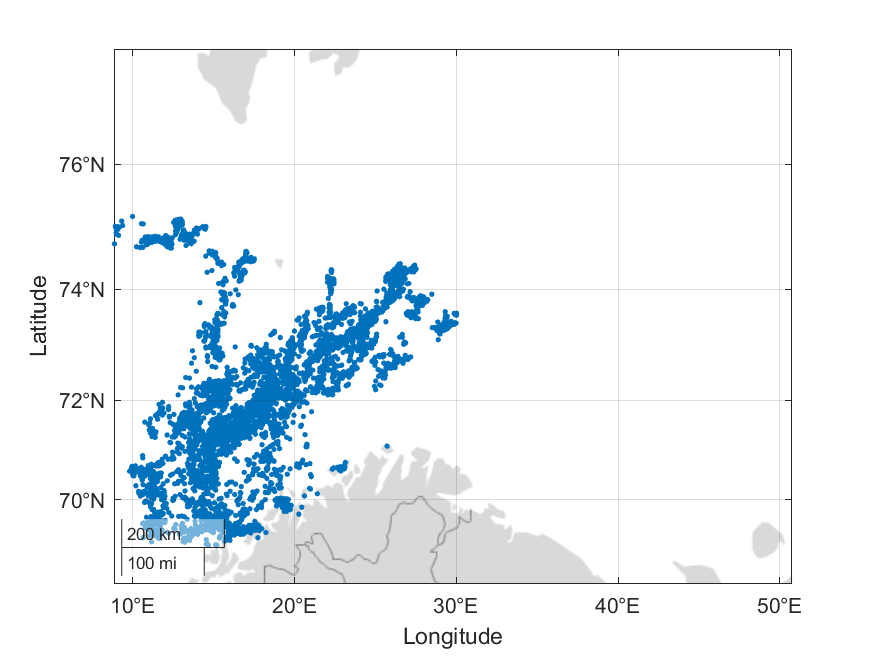} 
\caption{February 17} \label{fig:e2}
\end{subfigure}
\begin{subfigure}{0.33\textwidth}
\includegraphics[scale=0.375]{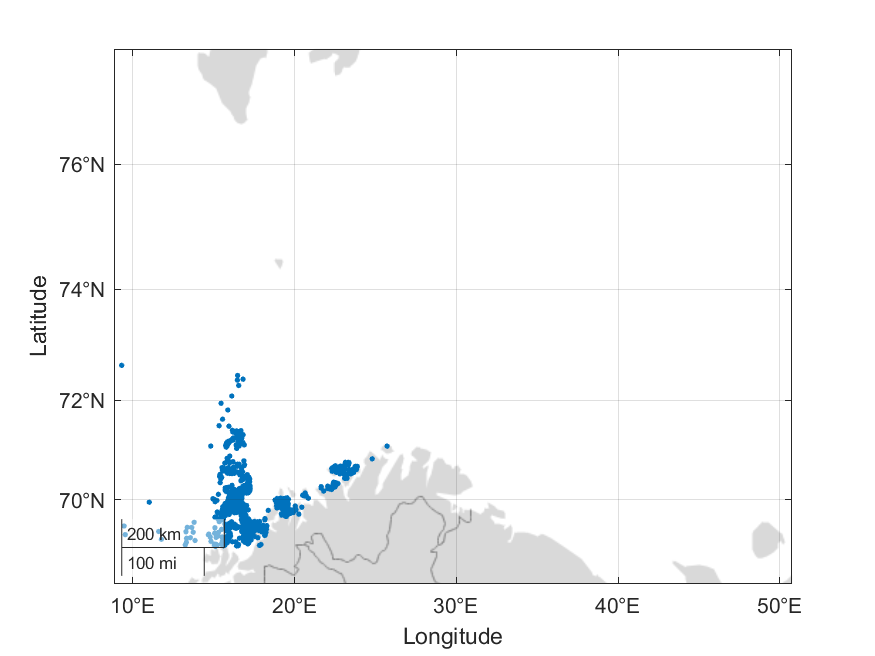}
\caption{March 31} \label{fig:f2}
\end{subfigure}
\medskip
\begin{subfigure}{0.33\textwidth}
\includegraphics[scale=0.375]{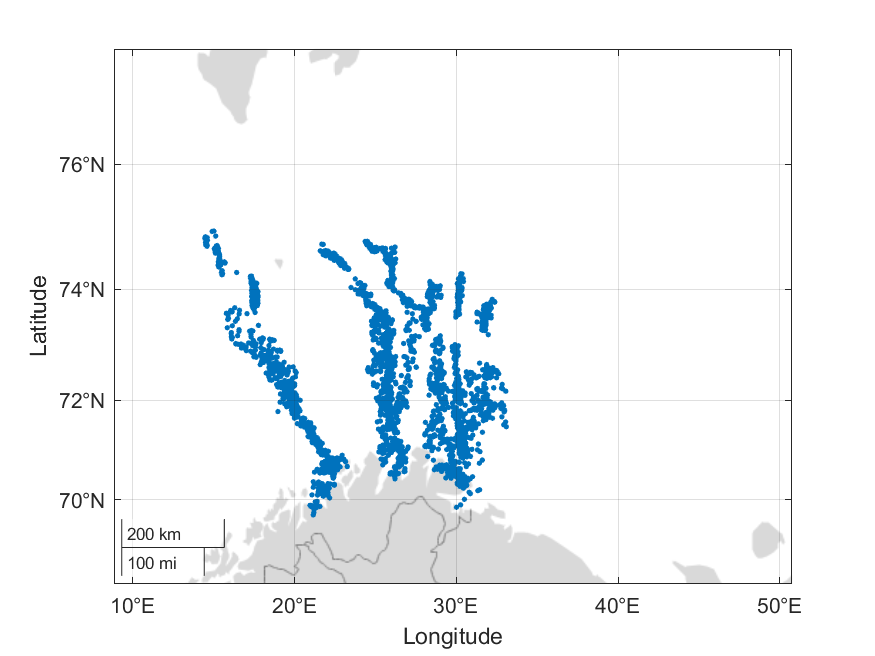} 
\caption{January  28} \label{fig:g2}
\end{subfigure}
\begin{subfigure}{0.33\textwidth}
\includegraphics[scale=0.375]{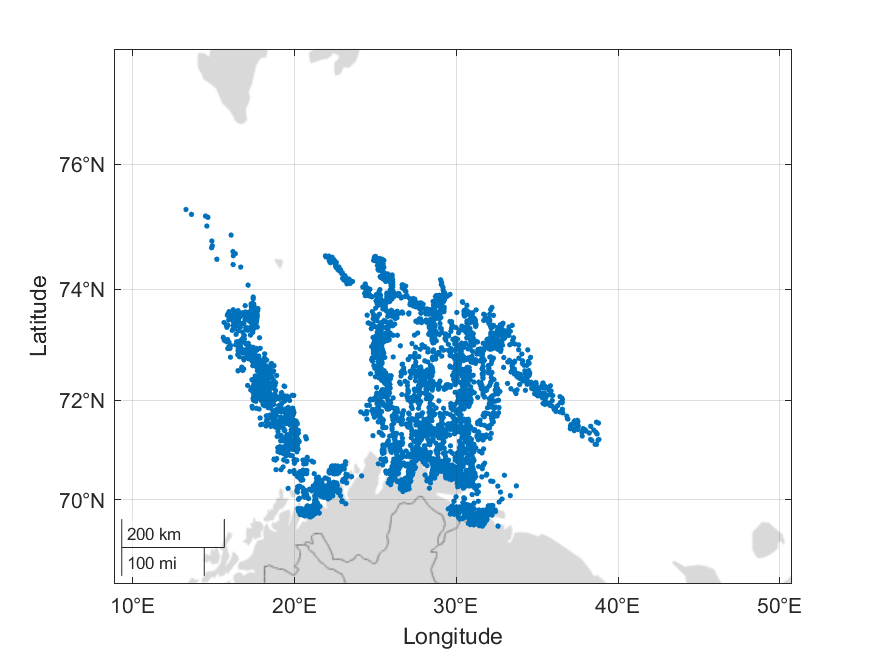} 
\caption{February 17} \label{fig:h2}
\end{subfigure}
\begin{subfigure}{0.33\textwidth}
\includegraphics[scale=0.375]{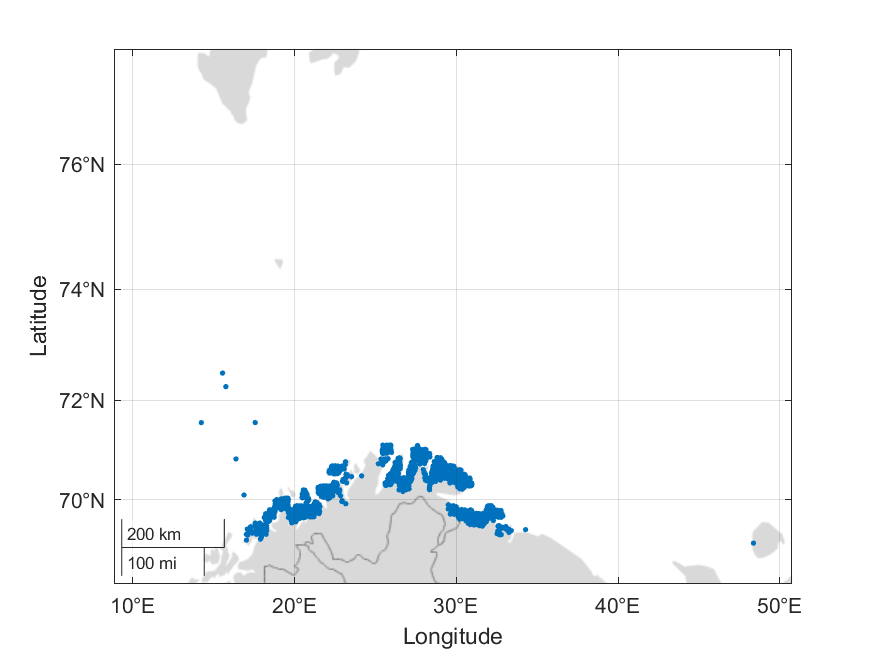} 
\caption{ March 31} \label{fig:i2}
\end{subfigure}
\caption{The distribution of the simulated capelin, represented by blue dots, is depicted when employing the adaptive migration model. The artificial neural networks (ANNs) were fine-tuned through the genetic algorithm (GA), where the fitness function was designed to maximize the sea surface temperature (SST) in the first set of figures (a-c) and in the second set of figures (d-f), a weighted combination of temperature and distance to spawning regions was considered.} \label{fig:fitness_compare}
\end{figure}

The test fitness function aimed to maximize the overall temperature, regardless of the proximity of the fish to the spawning regions. This comparison revealed that the migration trajectories became less realistic when only temperature was maximized. However, the migration trajectories improved in realism when we included the weighted distance to the spawning region as a component in the fitness function. The figures in Figure \ref{fig:fitness_compare} (a-c) illustrate the migration trajectories when only temperature is maximized. Conversely, when we integrated the weighted distance to the spawning region component into the fitness function, the migration trajectories became more realistic, as seen in Figure \ref{fig:fitness_compare} (d-f).

These sensitivity analyses offer valuable insights into the factors that impact capelin spawning migration simulations. They underscore the significance of initial distributions, releasing dates, and the choice of fitness function in faithfully representing capelin migratory behaviors. By comprehending these sensitivities, researchers can enhance their ability to interpret and analyze simulation results, thus making informed decisions when designing future experiments or modeling studies.
\printbibliography
\end{document}